\documentclass[11pt]{article}
\pdfoutput=1
\usepackage{authblk}
\usepackage[english]{babel}
\usepackage{amsmath,amssymb,exscale,color,graphicx}
\usepackage[utf8]{inputenc}
\usepackage{amsfonts}
\usepackage{amsthm}
\usepackage{mathrsfs}
\usepackage{xcolor}
\usepackage{caption}
\usepackage{subcaption}
\pdfoutput=1
\usepackage{dcolumn}
\usepackage{bm}

\usepackage{multirow,enumitem}
\usepackage{cite,color,url}
\usepackage[colorlinks=true
,urlcolor=blue
,anchorcolor=blue
,citecolor=blue
,filecolor=blue
,linkcolor=blue
,menucolor=blue
,linktocpage=true
,pdfproducer=medialab
,pdfa=true
]{hyperref}

\usepackage{slashed}
\usepackage{epsfig,psfrag,rotating,soul}
\usepackage{rotfloat}

\usepackage{hhline}
\usepackage{array}

\input epsf
\newcommand{\m}[1]{\marginpar{{\tiny *}} }

\evensidemargin \oddsidemargin
\allowdisplaybreaks

\begin{document}

\title{Double Higgs production at the HL-LHC: probing a loop-enhanced model with kinematical distributions}

\author[a]{Leandro~Da~Rold\thanks{daroldl@ib.edu.ar}}
\author[b]{Manuel~Epele\thanks{manuepele@fisica.unlp.edu.ar}}
\author[b]{Anibal~D.~Medina\thanks{anibal.medina@fisica.unlp.edu.ar}}
\author[b]{Nicol\'as~I.~Mileo\thanks{mileo@fisica.unlp.edu.ar}}
\author[b]{Alejandro Szynkman\thanks{szynkman@fisica.unlp.edu.ar}}
\affil[a]{Centro At\'omico Bariloche, Instituto Balseiro and CONICET, Av. Bustillo 9500, 8400, S. C. de Bariloche, Argentina}
\affil[b]{IFLP, CONICET - Dpto. de F\'{\i}sica, Universidad Nacional de La Plata, C.C. 67, 1900 La Plata, Argentina}

\maketitle

\begin{abstract}
We study di-Higgs production via gluon fusion at the high luminosity LHC in the presence of new physics, focusing on the $b\bar b\gamma\gamma$ final states. Taking a minimal set of three scalar leptoquarks (LQs) with cubic and quartic interactions with the Higgs and choosing four benchmark points with a light LQ, we perform a detailed analysis of differential distributions of the di-Higgs production cross section, studying the imprints of the new physics states running in the loops. Simulating the signal and main backgrounds, we study the influence of the new physics in differential distributions such as the invariant mass of the subsystems of final particles, the transverse momentum, and angular variables, finding in particular a resonance peak associated with the light LQ. It turns out that the angular separation of the photons, which is correlated with the resonance LQ peak, is a very sensitive observable that helps in discriminating the new physics signal from the Standard Model background. We find that for two of our benchmarks discovery could be reached with 3~ab$^{-1}$, whereas exclusion limits at 95\% C.L. could be claimed with 0.60-0.75~ab$^{-1}$. For the other two benchmarks that have heavier LQ masses significances of order 2$\sigma$ are possible for  3~ab$^{-1}$. A similar analysis could be applied to other loop-enhanced models.

\end{abstract}

\newpage

\tableofcontents

\section{Introduction}
\label{sec-introduction}
One of the most crucial measurements that can be accomplished in the high luminosity era at the LHC (HL-LHC) is the production of a Higgs ($h$) pair from the dominant gluon fusion production channel~\cite{Dicus:1987ic,Glover:1987nx,Plehn:1996wb,Dawson:1998py,Djouadi:1999rca,BarrientosBendezu:2001di}. In particular, this is a predicted process from the Standard Model (SM) of particle physics in which the quartic Higgs coupling $\lambda$ from the electroweak symmetry breaking (EWSB) potential enters explicitly~\cite{Baur:2002rb,Baur:2002qd,Baur:2003gpa,Baur:2003gp,Baglio:2012np}. Thus this measurement, along with the more difficult to asses triple $h$ production, may help in elucidating the mechanism of EWSB in the SM. It is well-known however that the gluon initiated double $h$ production in the SM suffers from a partial cancellation between the top quark dominated box and triangle diagrams, which becomes exact at threshold~\cite{Plehn:1996wb,Djouadi:1999rca,Baur:2002rb}. This leads to a cross section for double $h$ production that at leading order (LO) is $\sim 17$ fb$^{-1}$ and that at next-to-next-to-leading-order (NNLO) increases to $\sim 39$ fb$^{-1}$~\cite{LHCHiggsCrossSectionWorkingGroup:2016ypw}. Current measurements from ATLAS and CMS at a luminosity $\mathcal{L}\sim 140$ fb$^{-1}$ have little sensitivity, putting constraints to $\sigma_{hh}\lesssim (4\sim 8)\times \sigma_{hh,SM}$ for the pair of Higgs bosons decaying to $b\bar{b}\gamma\gamma$~\cite{ATLAS:2023gzn,CMS:2020tkr}. Considering the most promising Higgs decay combinations $b\bar{b}\gamma\gamma$, $b\bar{b}b\bar{b}$, $b\bar{b}\bar{\tau}\tau$, $b\bar{b}ZZ^{*}(4l)$ and $b\bar{b}WW^{*}(l\nu l\nu)$ , it is claimed that a 4.5$\sigma$ discovery significance in the SM is possible for $\mathcal{L}\sim 3$  ab$^{-1}$, when combining all channels and both experiments~\cite{ATLAS:2022hsp}, neglecting systematic uncertainties.

In this context, it becomes interesting to consider possible beyond the SM (BSM) contributions that could affect double $h$ production at the LHC~\footnote{Current estimations of the HL-LHC expect to reach a $\mathcal{L}=3$ ab$^{-1}$ by the year 2040.}, providing larger cross sections that may be probed at luminosities less than  3 ab$^{-1}$. Modifications to the quartic $\lambda$ as predicted by the SM have been explored, where usually large luminosities and center of mass energies ($\sqrt{s}\geq 14$ TeV) are required for discovery (see for example \cite{ATLAS:2017muo,Barr:2014sga,Kling:2016lay,Adhikary:2017jtu,Goncalves:2018qas}). New BSM particles running in the loops of the double $h$ gluon fusion diagrams are also an interesting alternative to enhance the production (loop-enhanced models), see for example the seminal Refs.~\cite{Plehn:1996wb,Belyaev:1999mx} for supersymmetric theories, \cite{Dib:2005re,Dawson:2012mk,Gillioz:2012se} for new fermions, \cite{Enkhbat:2013oba,Huang:2017nnw,Bhaskar:2022ygp} for LQs as well as other possibilities~\cite{Dolan:2012ac,Cao:2013si,Hespel:2014sla,Dawson:2015oha,Batell:2015koa}. In Ref.~\cite{DaRold:2021pgn} we considered a leptoquark (LQ) model with a minimal content of LQs and showed that at LO  $\sigma_{hh,LO}\sim 2.3\times \sigma_{hh,SM-LO}$ could be obtained for a light enough LQ, $m_{LQ}\sim 400$ GeV, via large cubic interactions in the LQ-Higgs potential, while at the same time avoiding all current experimental constraints, in particular: single Higgs measurements, LQ direct searches and electroweak precision measurements. 

In this work we do a full collider study for gluon-fused double $h$ production at the HL-LHC for four benchmark (BM) points of what we dubbed the ``scenario with a light LQ'' in our previous paper, focusing on the $b\bar{b}\gamma\gamma$ channel~\cite{Baur:2003gp}. We consider the main SM backgrounds for our signal and take our benchmarks to satisfy the latest measurements from single Higgs and LQ direct searches. We validate our Monte Carlo (MC) results against our previous numerical studies, finding excellent agreement. Studying the differential cross-section distributions, we find that the presence of the light LQ in the loop can provide a resonant enhancement that manifests itself clearly in the invariant di-Higgs mass $m_{hh}$ distribution but also in the separation between the two photons, $\Delta R_{\gamma\gamma}$, providing important handles that allow a strong background suppression. This is a unique property of the loop-enhanced models, which could serve as a clear distinction from BSM models that modify Higgs quartic interaction or via modifications from effective operators (EFTs)~\cite{Contino:2012xk,Goertz:2014qta,Azatov:2015oxa}.

The paper is organized as follows: in section~\ref{sec-NPmodel} we briefly recapitulate on the LQ model which enhances di-Higgs production via gluon fusion, its experimental constraints, and calculate the total and differential cross sections for the four BM points. In section~\ref{sec-simulations} we discuss how we simulate the BSM signal and SM backgrounds. In section~\ref{sec-analysis} we present our results, showing the power of considering the differential cross-section distributions. Finally, in section~\ref{sec-conclusions} we present our conclusions and briefly discuss the possibility of implementing machine learning tools for the analysis.

\section{LQ model for the enhancement of di-Higgs production}
\label{sec-NPmodel}
We consider a model with the following scalar LQs: $\tilde R_2\sim({\bf 3},{\bf 2})_{1/6}$, $S_1\sim(\bar{\bf 3},{\bf 1})_{1/3}$ and $\bar S_1\sim(\bar{\bf 3},{\bf 1})_{-2/3}$, allowing cubic interactions with the Higgs, that can lead to sizable corrections of di-higgs production at the LHC. The full Lagrangian of the model, as well as its phenomenology, have been discussed in Ref.~\cite{DaRold:2021pgn}. Below we describe the ingredients relevant for the study in this work. The most general renormalizable potential is
\begin{equation}
V=V_2+V_3+V_4
\label{eq-V}\\
\end{equation}
where
\begin{align}
V_2= & \, m_H^2|H|^2 + m_{\tilde R_2}^2|\tilde R_2|^2 + m_{S_1}^2|S_1|^2 + m_{\bar S_1}^2|\bar S_1|^2 \ ,
\label{eq-V2}\\[2mm]
V_3= &\,\mu_1 H^\dagger \tilde R_2 S_1 +\mu_2 \tilde R_2^j \epsilon_{jk} H^k\bar S_1 +{\rm h.c.} \ , \label{eq-V3}\\[2mm]
V_4=&\sum_{\phi} \lambda_\phi (\phi^\dagger\phi)^2+\lambda_1 H^\dagger H\tilde R_2^\dagger\tilde R_2+\lambda_2 H^\dagger HS_1^\dagger S_1+\lambda_3 H^\dagger H\bar S_1^\dagger\bar S_1 + \dots \ ,\label{eq-V4}
\end{align}
where the sum is over all the scalar fields: $\phi=H,\tilde R_2,S_1,\bar S1$, the dots stand for quartic terms that do not involve the Higgs and $\mu_{1,2}$ are the dimension-full cubic couplings.

To avoid proton decay, we assume that baryon ($B$) and lepton ($L$) numbers are conserved. We take $\mu_1=0$ and assign the following numbers to the LQs:
$B(\tilde R_2)=-B(\bar S_1)=2B(S_1)=-2/3$, $L(\tilde R_2)=L(\bar S_1)=0$ and $L(S_1)=-1$. We work in the ``scenario with a light LQ'' (see Ref.~\cite{DaRold:2021pgn} for details).

The Higgs vacuum expectation value (vev), as well as the cubic coupling $\mu_2$ mix the LQs of equal electric charge. The mass matrices can be diagonalized by unitary transformations~\cite{DaRold:2021pgn}, such that in the mass basis, the potential reads: 
\begin{equation}
V=\sum_{i,q}|\chi^q_i|^2m_{\chi^q_i}^2 + h \sum_{i,j,q}\chi^{q*}_i\chi^q_j \mathscr{C}^{(q)}_{ij} + h^2 \sum_{i,j,q}\chi^{q*}_i\chi^q_j {\cal Q}^{(q)}_{ij} + \dots
\label{eq-Vmass}
\end{equation}
with $\mathscr{C}^{(q)}$ and ${\cal Q}^{(q)}$ the cubic and quartic couplings, respectively, $i,j=1,2$ and $q=u,d$, indicating the electromagnetic charge of the states, $u$ for charge $2/3$ and $d$ for $-1/3$. In Ref.~\cite{DaRold:2021pgn} one can find the mass matrices, as well as the matrices of the cubic and quartic interactions, in the gauge basis of Eq.~(\ref{eq-V}). 

There are two sources for cubic couplings: the coefficients $\mu_i$, that although in the gauge basis only couple different LQs, in the mass basis lead also to diagonal components proportional to the mixing angles, and the quartic coefficients $\lambda_i$ with one Higgs vev, that in the mass basis lead to diagonal as well as off-diagonal interactions. The quartic couplings are proportional to $\lambda_{i}$, and in the mass basis lead to diagonal and off-diagonal interactions.

For the study of double Higgs production, we have made a random scan of several hundred points and selected four benchmark points (BM) that span a range of masses and couplings of the LQs. The corresponding parameters of the potential of Eq.~(\ref{eq-V}) are in the following ranges: $m_{LQ}\in(0.6,3.4)$~TeV, $\mu_2\in(0.9,5.7)$~TeV, $\lambda_i\sim{\cal O}(1)$ and $\lambda_1'=0$.~\footnote{The $T$ parameter requires a moderate value of $\lambda_1'$~\cite{DaRold:2021pgn}, for simplicity we take it vanishing in this work.} With $\mu_1=\lambda_1'=0$ there is no mixing between down-type LQs. Since the di-Higgs production cross-section is mostly driven by the mass and couplings of the lightest LQ, we choose BM points in which the lightest mass ranges from $\sim$ 400 to 800 GeV and that have sizable couplings. The masses and couplings of the BMs in the mass basis are shown in table~\ref{t-mass}.~\footnote{The coupling $\mathscr{C}^{(u)}_{22}$ of BM4 is accidentally suppressed, compared with the other BMs. However, we have analyzed other points with $m_{\chi^u_2}\simeq 800$~GeV and $\mathscr{C}^{(u)}_{22}\sim 10^3$~GeV, obtaining results very similar to those presented along this article for BM4.} The parameters of the potential of Eq.~(\ref{eq-V}) that generate these BMs can be computed analytically by performing the inverse unitary rotations that lead to Eq.~(\ref{eq-Vmass}). 

\begin{table}[ht]
\renewcommand*{\arraystretch}{1.8}
\small
\begin{center}
\begin{tabular}{|c||c|c|c|c|}
\hline
{\bf Quantity} & \rm{BM}1 & \rm{BM}2 & \rm{BM}3 & \rm{BM}4
\\ \hhline{|=||=|=|=|=|}
$m_{\chi^u_1}$[GeV] & 2297 & 2325 & 2054 & 2524 
\\ \hline
$m_{\chi^u_2}$[GeV] & 464 & 512 & 621 & 800
\\ \hline
$m_{\chi^d_1}$[GeV] & 2250 & 2269 & 1998 & 3409 
\\ \hline
$m_{\chi^d_2}$[GeV] & 899 & 938 & 1708 & 2497 
\\ \hline 
\rule{0pt}{2.3\normalbaselineskip}  $\mathscr{C}^{(u)}$[GeV] & {\footnotesize $\begin{pmatrix} 2068 & -3650 \\ -3650 & -1297\end{pmatrix}$} & {\footnotesize$\left(\begin{array}{cc} 2595 & 3539 \\ 3539 & -1530 \end{array}\right)$} & {\footnotesize$\left(\begin{array}{cc} 1932 & 3124 \\ 3124 & -1541 \end{array}\right)$} & {\footnotesize$\left(\begin{array}{cc} 1582 & -2785 \\ -2785 & 92 \end{array}\right)$}
\\[6mm] 
\hline
$\mathscr{C}^{(d)}$[GeV] & diag(412,1213) & diag(597,1182) & diag(191,1087) & diag(968,507)
\\ \hline
\rule{0pt}{2.3\normalbaselineskip} ${\cal Q}^{(u)}$ & {\footnotesize$\left(\begin{array}{cc} 1.013 & 0.098 \\ 0.098 & 0.552 \end{array}\right)$} & {\footnotesize$\left(\begin{array}{cc} 2.064 & -0.414 \\ -0.414 & 0.099 \end{array}\right)$} & {\footnotesize$\left(\begin{array}{cc} 0.596 & -0.103 \\  -0.103 & 0.197 \end{array}\right)$} & {\footnotesize$\left(\begin{array}{cc} 1.764 & 0.016 \\ 0.016 & 1.638 \end{array}\right)$}
\\[6mm] \hline
${\cal Q}^{(d)}$ & diag(0.838, 2.465) & diag(1.213,2.403) & diag(0.389,2.209) & diag(1.968,1.030)
\\ \hline
\end{tabular}
\end{center}
\caption{Masses, cubic $\mathscr{C}^{(q)}$ and quartic ${\cal Q}^{(q)}$ couplings of the LQs corresponding to the BM points, in the mass basis Eq.~(\ref{eq-Vmass}).}
\label{t-mass}
\end{table}

It is worth stressing that none of the selected BM points is excluded by theoretical or experimental constraints related to electroweak precision tests, flavor, or collider physics. In particular, the impact of LQs on oblique corrections, the $Zb\bar b$ coupling, and flavor-changing transitions has been studied in Ref.~\cite{DaRold:2021pgn}. Among them, it was shown there that the $T$-parameter imposes restrictive constraints on the model and, as commented above, specifically on the coupling $\lambda_1'$. Without overlooking the fact that the effect of this coupling on double Higgs production is mild, the simple choice $\lambda_1'=0$ in all the BM points alleviates these constraints. The LQs, being colored and electrically charged fields, correct to one-loop order the Higgs interactions with gluons and photons. This affects the single Higgs production via gluon fusion and the Higgs decay into a pair of photons. The corrections to these couplings can be studied in terms of the $\kappa$ formalism~\cite{LHCHiggsCrossSectionWorkingGroup:2013rie}. We have checked that for the BM points $\delta\kappa_g$ and $\delta\kappa_\gamma$ are compatible at $2\sigma$ level with a combined fit performed by ATLAS including analyses with luminosities ranging from 24.5 fb$^{-1}$ to 139 fb$^{-1}$~\cite{ATLAS:2020qdt}, with BM1, BM2 and BM3 being between $1\sigma$ and $2\sigma$, whereas BM4 has a very small correction compared with the SM (see also Ref.~\cite{CMS:2018uag} by CMS for less restrictive limits).~\footnote{The contours in the $(\delta\kappa_g,\delta\kappa_\gamma)$ plane at 68\% and 95\% C.L. are obtained in Ref.~\cite{ATLAS:2020qdt} from a combined fit with two degrees of freedom where all other couplings are fixed to their SM values and the invisible or undetected Higgs boson decays do not contribute. In a previous analysis by ATLAS~\cite{ATLAS:2019nkf} a combined fit with two degrees of freedom is performed too. In this case, our four benchmarks are also compatible with the experimental limits at  $2\sigma$ level. In a more recent analysis, ATLAS made a combined fit with $\kappa_{Z\gamma}$ profiled which corresponds to three degrees of freedom (see auxiliary material in Ref.~\cite{ATLAS:2022vkf}). In this case, BM1 and BM2 might be in tension. However, since we do not have enough information on the way the limits are projected as contours onto the $(\delta\kappa_g,\delta\kappa_\gamma)$ plane, we avoid any conclusive statement in this regard.} In our analysis we will also take into account this effect in the di-photon decay of the Higgs. 

The lightest LQ in each BM point decays dominantly to dijets. Important constraints arise in this case from the following searches. Reinterpreting the experimental bounds on the mass of stops decaying promptly through R-parity-violating couplings as limits on LQ masses, direct searches of nonresonant pair production of resonances lead to relevant restrictions on the allowed LQ spectrum. In an analysis performed by ATLAS at 36.7 fb$^{-1}$~\cite{ATLAS:2017jnp} stop masses are excluded from 100 GeV to 410 GeV for stop decays into two light quarks and in the ranges $(100-470)$ GeV and $(480-610)$ GeV for decays into $b$ and a light quark. Using 35.9 fb$^{-1}$ CMS~\cite{CMS:2018mts} excludes masses in the range $(80-520)$ GeV for final states with two light quarks, and reports the exclusion of values $(80-270)$ GeV, $(285-340)$ GeV and $(400-525)$ GeV for a $b$ and a light quark in the final state. In a search carried out by CMS at 138 fb$^{-1}$~\cite{CMS:2022usq} stop masses are excluded in the range $(580-770)$ GeV for stops decaying into a pair of light quarks. In light of these results, the mass of the lightest LQ in BM points 1, 2, and 3, as shown in table~\ref{t-mass}, seems to be excluded. However, it is worthwhile to remark that in all the referenced searches it is assumed that the stop decays into the selected final state with a branching ratio (BR) equal to 1 and, therefore, any relaxation of this assumption might lead to less restrictive constraints. For the purpose of this study it is important to point out that the LQs interactions with fermions enter as higher order corrections to Higgs productions. In this sense, it is implicitly understood that in the four BM points the LQ-fermion couplings result in a proportion of branching ratios that fulfill the experimental limits. This is not difficult to achieve by assuming, for instance, a BR equal to 0.7 for decays into light quarks and a BR equal to 0.3 for decays into $b$ and a light quark. The next to lightest LQ in each BM point decays dominantly into quarks and leptons. There are tight constraints coming from direct searches of LQs in this case as well. Bounds on LQs decaying predominantly into third generation fermions can be found in Refs.~\cite{CMS:2020wzx,ATLAS:2021oiz,ATLAS:2021yij,ATLAS:2023uox}, with a bound of $\sim 0.8-1.5$~TeV, depending on the BRs. For decays into quarks of the third generation and light leptons, the bounds are $\sim 1-1.4$~TeV~\cite{ATLAS:2022wcu}, according to different values of the BRs. The most stringent bounds for decays into light fermions are $0.8-1.8$~TeV considering BRs as low as 0.1~\cite{ATLAS:2020dsk}. In the case of decays into light quarks and $\tau$ Ref.~\cite{CMS:2023bdh} has considered single production that depends on the Yukawa couplings obtaining bounds $\sim 0.6-2$~TeV. Since these constraints depend on the BRs, we follow the same assumption made above for the lightest LQ decaying into dijets: LQs couplings are such that the branching ratios lead to decays compatible with the experimental limits. To accomplish this restriction is easier than in the case of the lightest LQ since now LQ couplings beyond those to quarks and leptons are present and then more decay channels are available. Finally, as mentioned, these analyses from ATLAS and CMS assume prompt decays. We have analyzed in Ref.~\cite{DaRold:2021pgn} the experimental limits on long-lived LQs. We have now proceeded as in that work and assumed that the LQ-fermion couplings are small enough to avoid flavor issues but also sufficiently large to evade long-lived particle searches.

An interesting observation may be the following. At least in recent experimental searches of colored particles, as Ref.~\cite{CMS:2022usq}, the minimum explored mass corresponds to 500 GeV. The reason behind this would lie perhaps in some degree of confidence that lower masses are already excluded. As we discussed above, the exclusion limits are compatible only with the assumption that a given final state has a branching ratio equal to 1. Thus, if future searches focus on large values of masses, it could be possible that the first signal of light LQs appears in single Higgs processes through corrections to $\kappa_g$ and $\kappa_\gamma$ since LQ contributions to these quantities are potentially large. For instance, we can easily see this in BM1 and BM2 where $(\delta\kappa_g,\delta\kappa_\gamma)$ is close to the $2\sigma$ limit with values $(-0.121,0.045)$ and $(-0.116, 0.043)$, respectively~\cite{ATLAS:2020qdt}. Although to a lesser extent, BM3 is still above the $1\sigma$ level with $(\delta\kappa_g,\delta\kappa_\gamma)=(-0.093,0.029)$.  

As thoroughly discussed in Ref.~\cite{DaRold:2021pgn}, in the present model there are correlations between single and double Higgs production at hadron colliders, since the same colored particles are involved in the virtual processes. However both production channels are complementary, at least in the following way: on the one hand, single Higgs is sensitive to corrections to the Wilson coefficient of effective operators as $hG^2$, but it is difficult to identify the source of those corrections~\footnote{For an alternative in single Higgs see Ref.~\cite{Grojean:2013nya} where a high transverse-momentum jet is required.}; on the other hand, di-Higgs production is sensitive to the properties of the virtual particles via the kinematical distributions of the di-Higgs system, as we will show in the next sections. Though precise measurements of single Higgs production alone could discard the BM points of the model, the discovery of new physics in single Higgs would require more information to decipher its nature, part of which could be provided by di-Higgs production. Furthermore, there are couplings that mix different LQs that enter in di-Higgs and are absent in single Higgs production.

\subsection{Di-Higgs production}
\begin{figure}
    \centering
    \includegraphics[width=.3\textwidth]{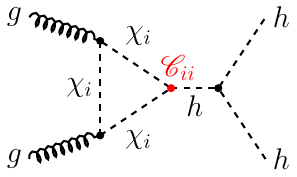}
    \includegraphics[width=.3\textwidth]{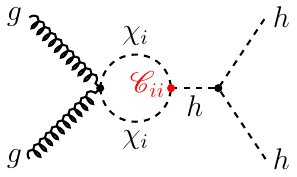}
    \includegraphics[width=.3\textwidth]{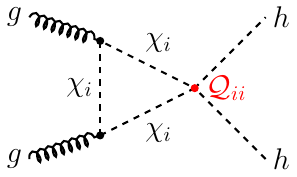}

    \vspace{.75cm}
    \includegraphics[width=.3\textwidth]{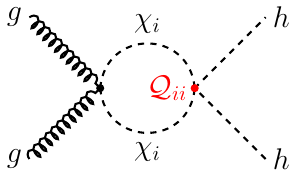}
    \includegraphics[width=.3\textwidth]{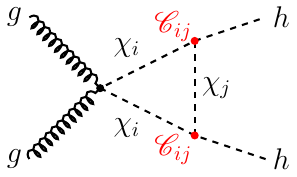}
    \includegraphics[width=.3\textwidth]{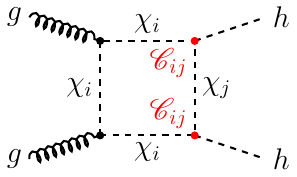}
    \caption{Diagrams mediated by scalar leptoquarks that contribute to di-Higgs production.}
    \label{fig-feyn-diags}
\end{figure}

The di-Higgs production cross-section in the presence of LQs has been computed in Refs.~\cite{DaRold:2021pgn,Enkhbat:2013oba}, in this article we follow the calculation of Ref.~\cite{DaRold:2021pgn}. The di-Higgs cross section via gluon fusion $\sigma_{hh}$ is driven by the cubic and quartic couplings of the potential that involve the Higgs field. The cubic couplings enter in the Feynman diagrams of di-higgs production in several ways, see Fig. \ref{fig-feyn-diags}: the triangle is linear in the diagonal components $\mathscr{C}^{(q)}_{ii}$, whereas the box is quadratic in $\mathscr{C}^{(q)}$, with $\mathscr{C}^{(q)}_{ii}$ entering in diagrams with one LQ running in the loop and $\mathscr{C}^{(q)}_{i\neq j}$ in diagrams with two different LQs in the loop. For quartic couplings only the diagonal components play a role, ${\cal Q}^{(q)}_{ii}$, entering in the diagrams with two Higgs fields attached to the same vertex of the loop.
 
We wrote our own numerical program to calculate the differential di-Higgs production cross-section at LO. The software implements {\tt LHAPDF6}~ \cite{Buckley_2015} to evaluate the gluon PDF of the proton at the center-of-mass energy scale of the scattering process. We chose the set PDF4LHC15\_nnlo\_mc~\cite{Butterworth_2016} and used {\tt LoopTools}~\cite{Hahn:1998yk} to numerically compute the scalar one-loop integrals involved in the SM and new physics form factors. We show the results in table~\ref{t-xs}, normalized to the SM one.

\begin{figure}[t!]
\begin{center}
\includegraphics[width=0.475\textwidth]{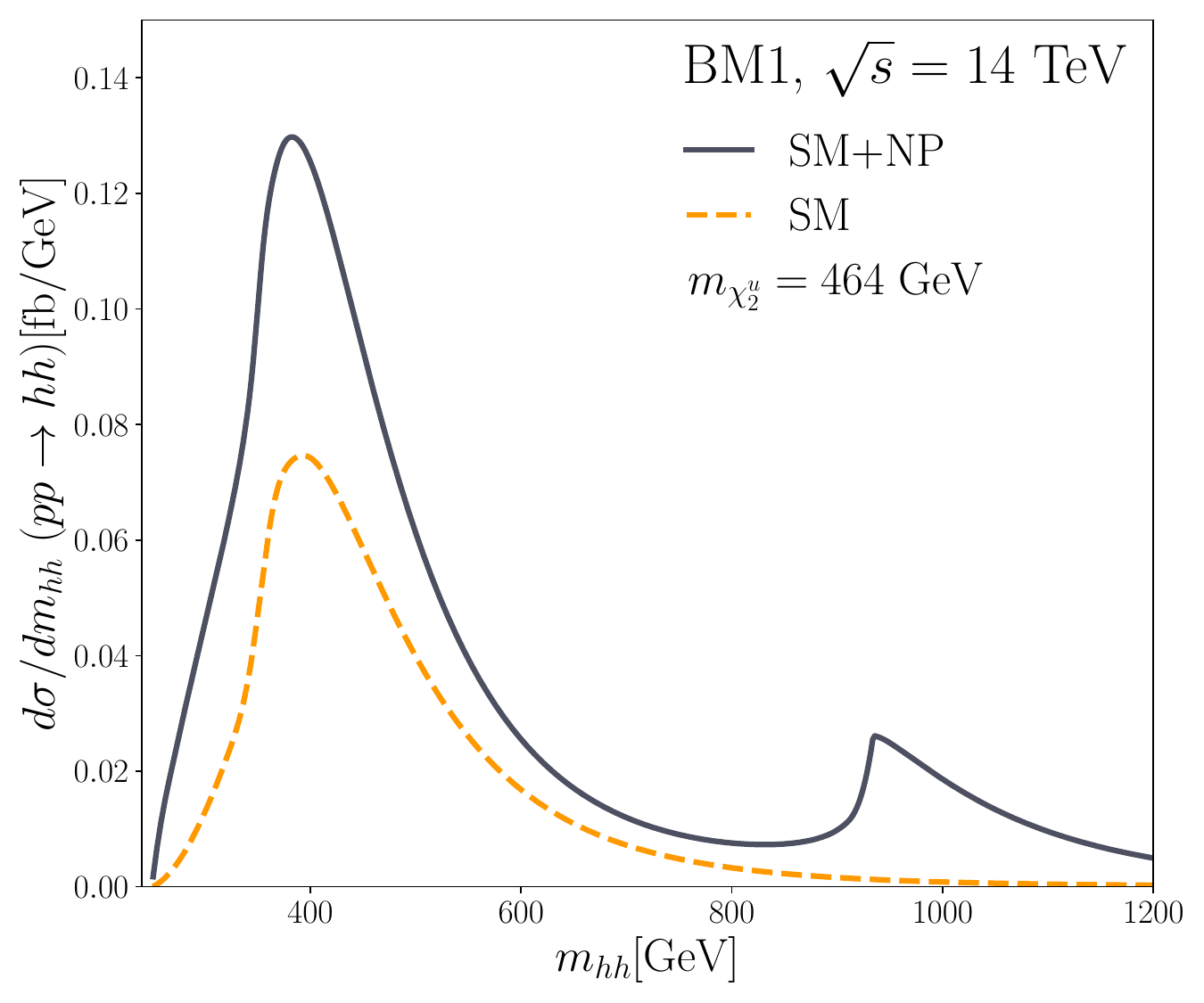}
\includegraphics[width=0.475\textwidth]{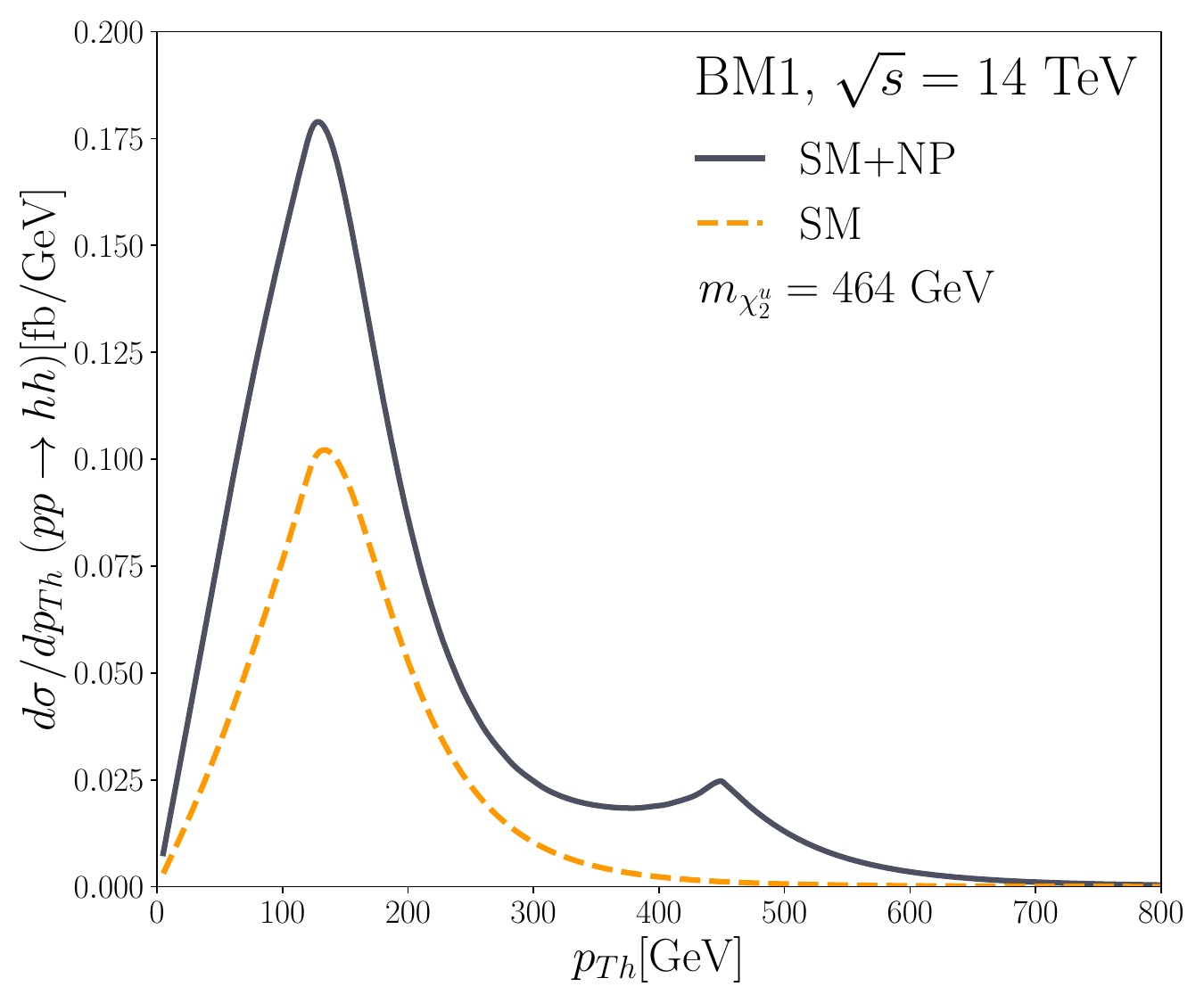}

\includegraphics[width=0.475\textwidth]{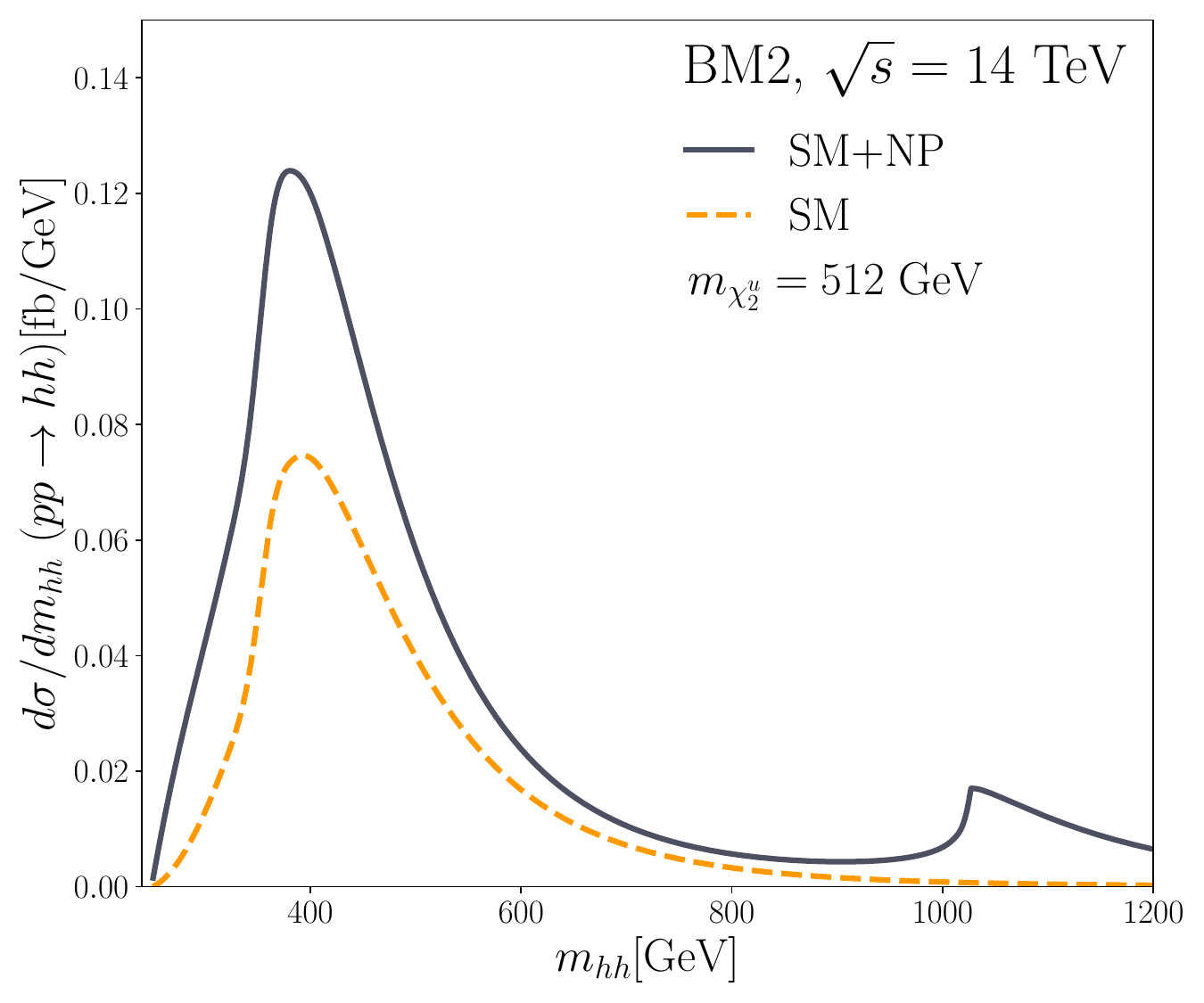}
\includegraphics[width=0.475\textwidth]{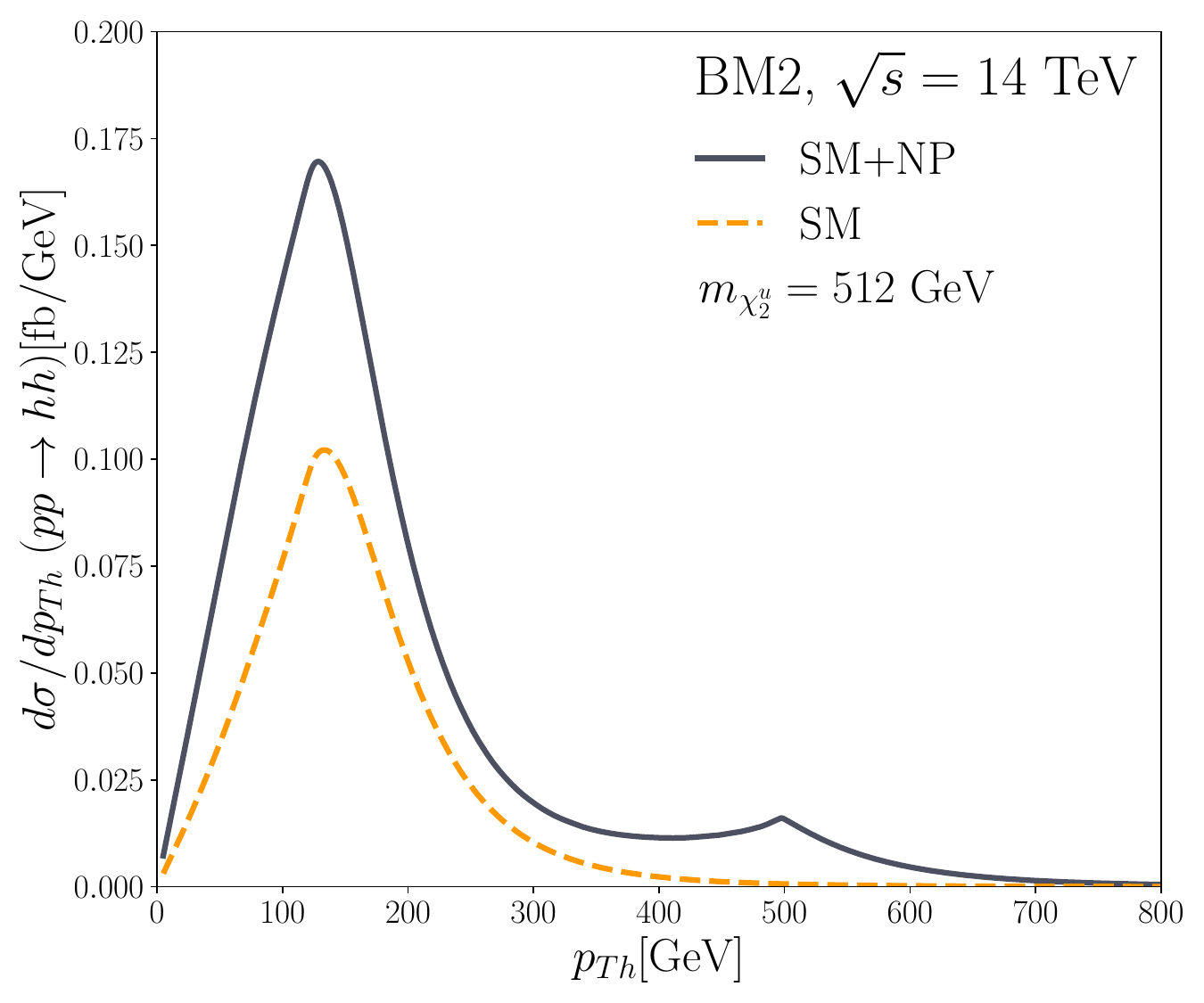}
\caption{Figures from theoretical calculations for benchmarks BM1 and BM2, at $\sqrt{s}=14$~TeV.}
\label{fig-mhh-pT12}
\end{center}
\end{figure}

\begin{figure}[t!]
\begin{center}
\includegraphics[width=0.475\textwidth]{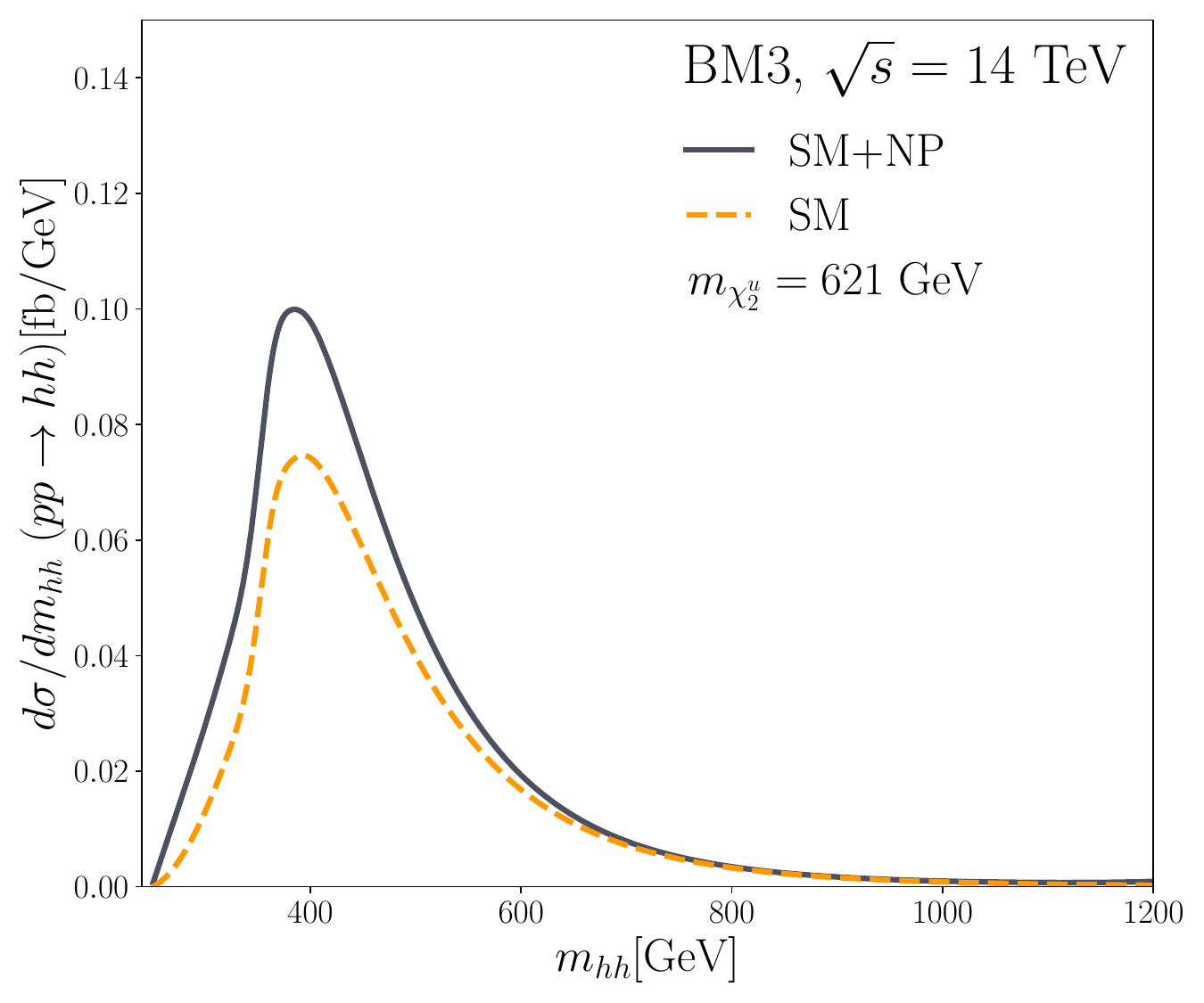}
\includegraphics[width=0.475\textwidth]{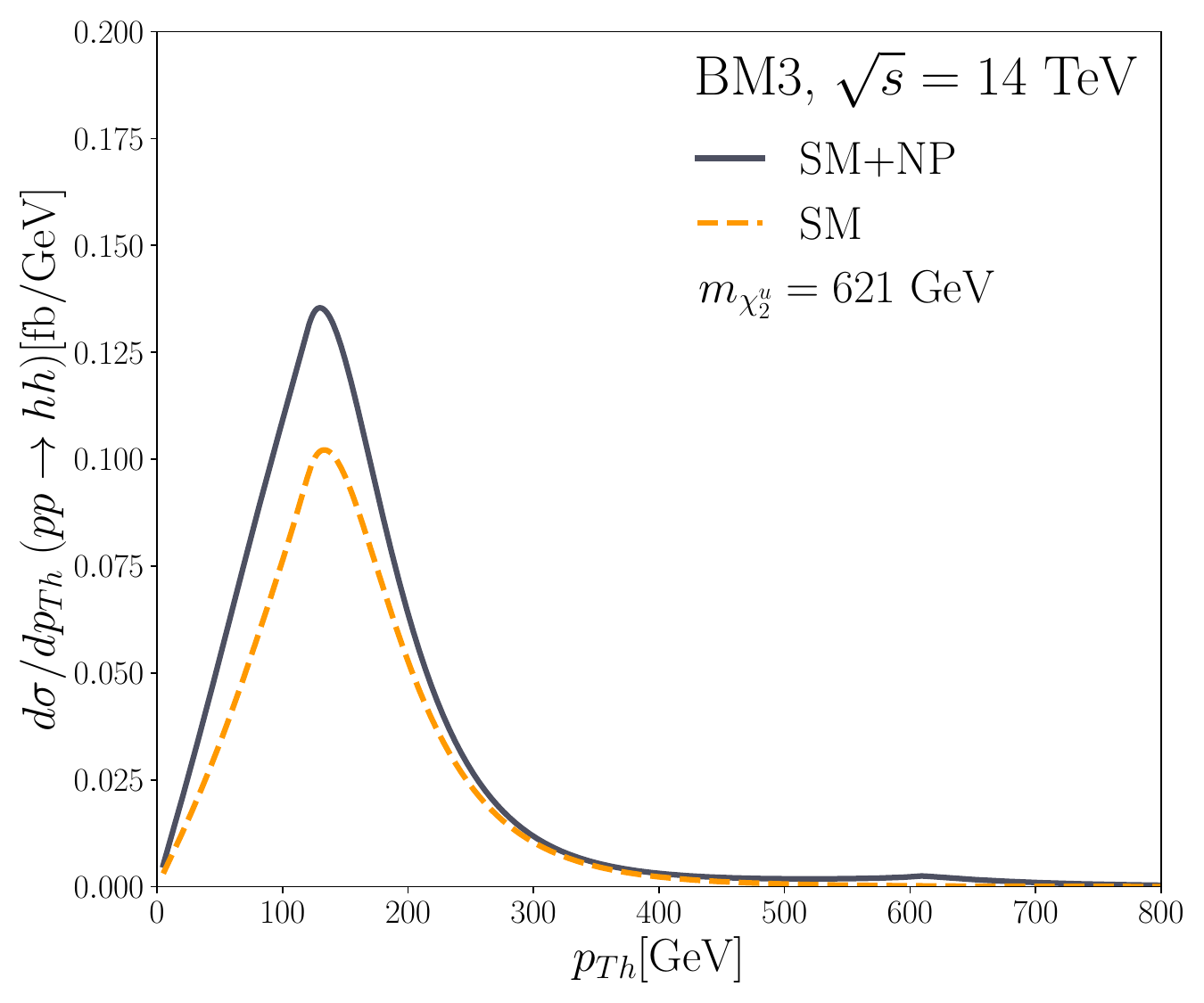}

\includegraphics[width=0.475\textwidth]{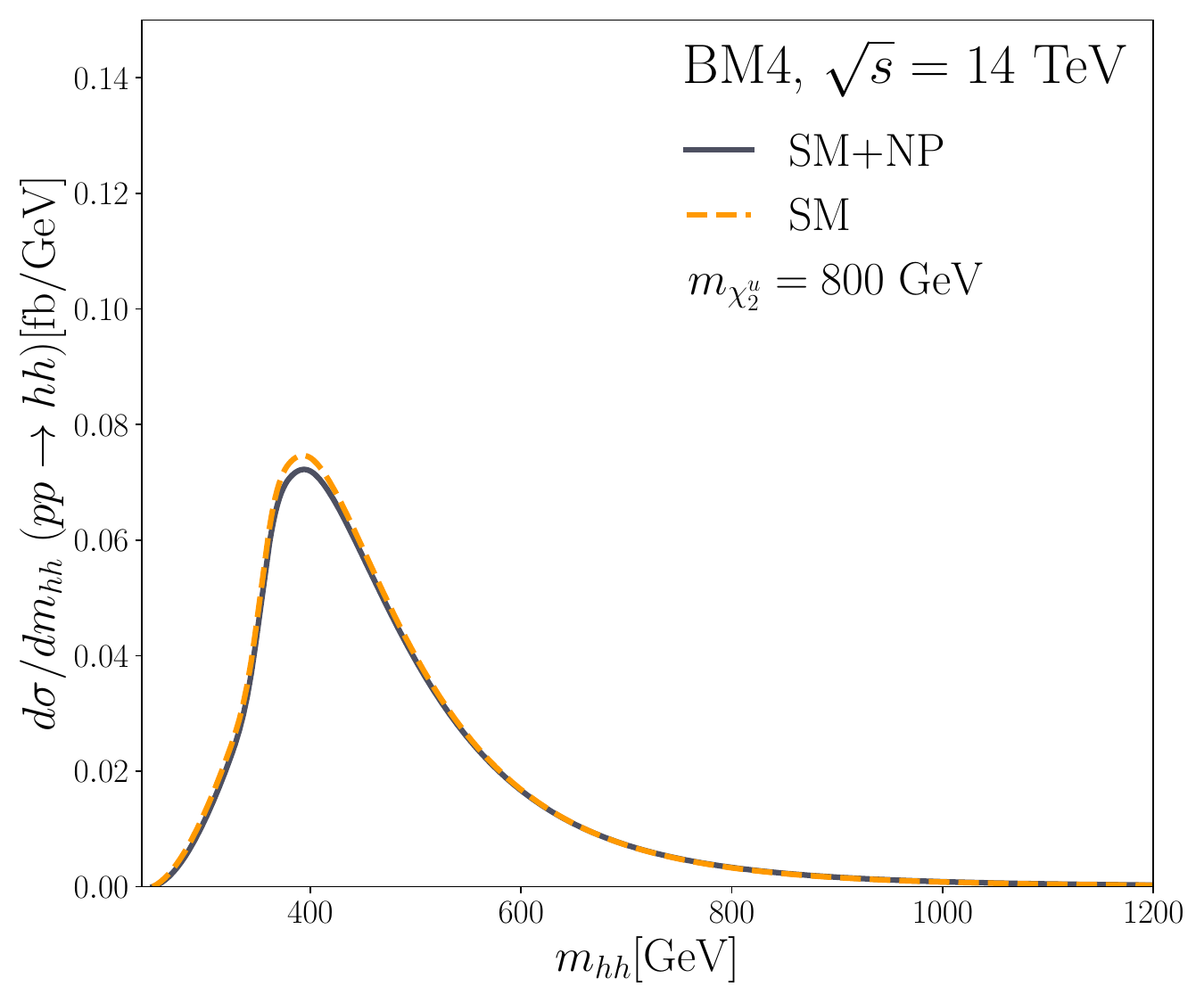}
\includegraphics[width=0.475\textwidth]{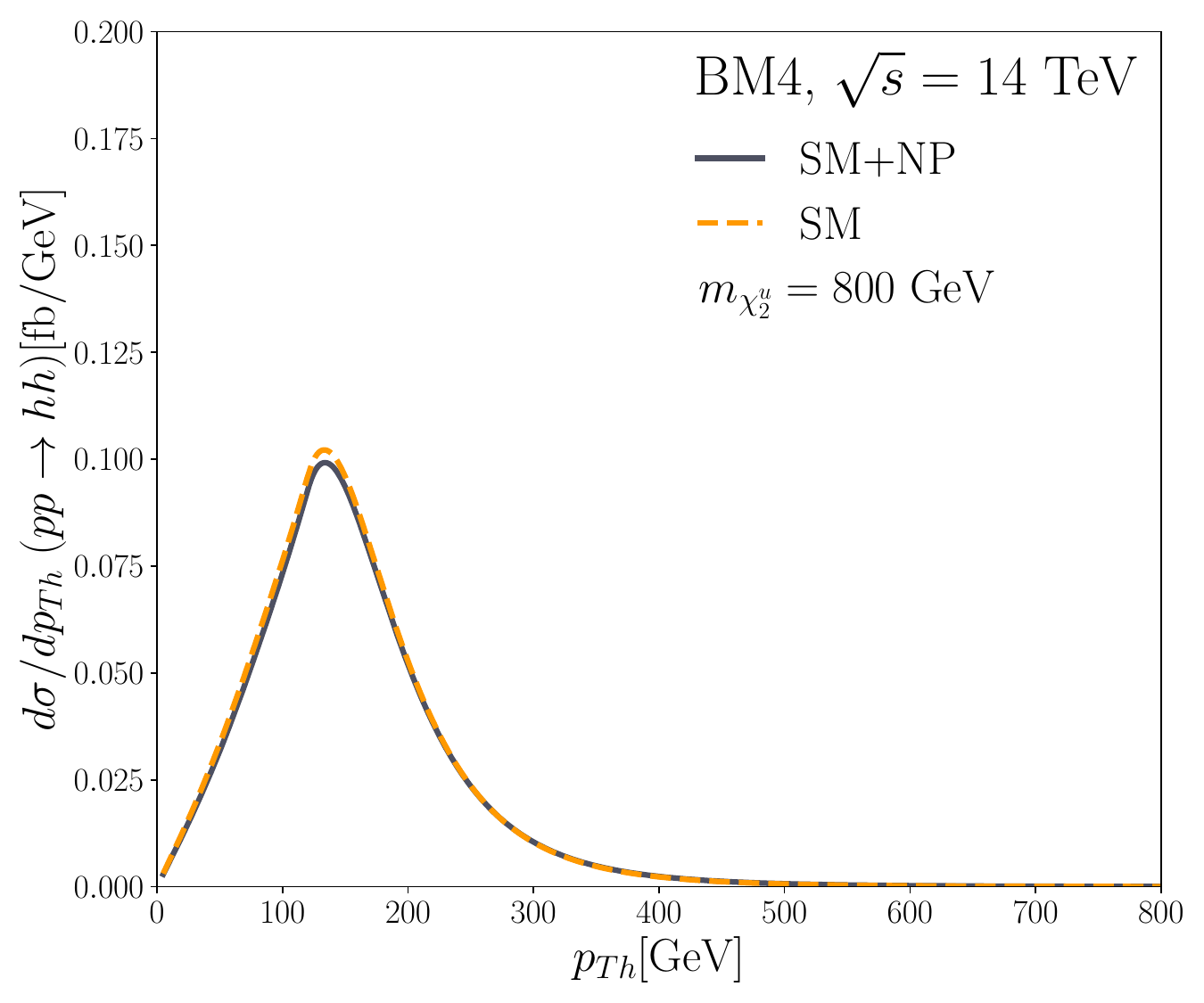}

\caption{Figures from theoretical calculations for benchmarks BM3 and BM4, at $\sqrt{s}=14$~TeV.}
\label{fig-mhh-pT34}
\end{center}
\end{figure}

In Figs.~\ref{fig-mhh-pT12} and~\ref{fig-mhh-pT34} we show the LO prediction of the differential cross sections for di-Higgs production as a function of the invariant di-Higgs mass $m_{hh}$ and transverse momentum of one of the Higgses $p_{Th}$, at the $\sqrt{s}=14$ TeV LHC, for the four chosen BM points in solid black and the SM prediction in dashed orange. Both the BM and the SM predictions were calculated using the same numerical code from our previous work~\cite{DaRold:2021pgn}. The influence of the light LQ on the differential distribution which shows up as a resonant peak can clearly be appreciated in both distributions ($m_{hh}$ and $p_{Th}$) for BM1 and BM2 in Fig.~\ref{fig-mhh-pT12}, whereas the peak is barely perceptible for BM3 and totally imperceptible for BM4 as can be appreciated in Fig.~\ref{fig-mhh-pT34}, where the LQ mass is heavier. This resonant peak is a feature of the loop-enhanced models and in the invariant mass distribution it peaks at $m_{hh}\sim 2m_{LQ}$, a region in which, for the masses considered, one may expect small SM backgrounds. Notice that the light LQ not only produces the resonance but also affects the peak associated with the top quark running in the loop at $m_{hh}\sim 2m_{t}$ (similarly for $p_{Th}$). It is useful to compare with the SM curve in which the top quark dominates the loop contribution. For BM1, BM2, and to a lesser degree BM3, we see a clear enhancement on the first peak of the distributions when comparing the solid black with the dashed orange curve (reaching roughly two times the SM peak value for BM1 and BM2). This is a consequence of the interference between the top quark (SM) and the LQ (BSM) in the di-Higgs loop production. Another interesting feature related to this is the observation that more Higgs pairs are expected with larger $p_T$ than in the pure SM case. Thus these Higgs bosons will tend to be boosted, implying that their decay products are collimated. This important observation will be exploited in the coming sections and shows to be a powerful handle in suppressing SM background. Note that for BM3 only a slight enhancement in the top quark peak is observed while for BM4 we obtain roughly the SM prediction (with a tiny suppression on the peak).

\begin{table}
\renewcommand*{\arraystretch}{1.5}
\begin{center}
\begin{tabular}{|c|c|c|c|c|}
\hline
$\sigma_{hh}$ & \rm{BM}1 & \rm{BM}2 & \rm{BM}3 & \rm{BM}4 
\\ \hhline{|=|=|=|=|=|}
$(\sigma_{hh}/\sigma^{\rm SM}_{hh})^{\rm LO}$ & 2.16 & 1.96 & 1.37 & 0.98  \\ 
\hline
\end{tabular}
\caption{Di-Higgs production cross-section for the BMs, normalized with respect to the SM one at LO.}
\label{t-xs}
\end{center}
\end{table}

\section{Simulations}
\label{sec-simulations}
To study the di-Higgs generation at the level of final states and reconstructed events after detection, we have simulated the signal and backgrounds with {\tt MadGraph5\_aMC@NLO} (MG5)~\cite{Alwall:2011uj,Alwall:2014hca} with the PDF set PDF4LHC15\_nnlo\_mc~\cite{Butterworth_2016}, at $\sqrt{s}=14~\rm{TeV}$. We have considered up to one extra jet with the parton-jet MLM matching scheme~\cite{Mangano:2006rw}. For the simulation of the Higgs decays, we have used {\tt MadSpin}~\cite{Artoisenet:2012st}. Parton showering and hadronization have been implemented with {\tt Pythia~8}~\cite{Sjostrand:2006za,Sjostrand:2007gs,Sjostrand:2014zea}, and a fast detector simulation has been carried out with {\tt Delphes 3}~\cite{deFavereau:2013fsa}, using the default ATLAS card for the high luminosity LHC.

\subsection{Signal}
To be able to simulate signal events, that require one-loop calculations, we have implemented the model of Eq.~(\ref{eq-V}), with LQs, in {\tt FeynRules}~\cite{Degrande:2011ua,Alloul:2013bka} at tree level. By making use of {\tt FeynArts}~\cite{Hahn:2000kx} and {\tt NLOCT}~\cite{Degrande:2014vpa} we have generated the new Lagrangian renormalized at one loop level and exported the corresponding model in UFO format~\cite{Darme:2023jdn} ready for simulations with MG5. We have checked that the simulations with this model correctly reproduce the SM cross-section and distributions by comparison with the implementation of the SM at one loop already included in MG5, namely the loop\_sm. We have also checked that in the presence of LQs the simulations reproduce the cross-sections and ratios shown in table~\ref{t-xs} as well as the distributions described in the previous section.

The UFO of the model is available in \url{https://github.com/manuepele/SM_LQs.git}. The external parameters of the model are the physical masses of the LQs and the quartic couplings in the gauge basis, the couplings in the mass basis are internally computed.

Armed with these tools we have simulated the di-Higgs production adding an extra jet with the aim of obtaining more realistic differential distributions of the cross section due to the large hadronic activity. The parton-jet matching has been performed by using the MLM scheme. In order to normalize the production cross section we consider the SM value of  Ref.~\cite{LHCHiggsCrossSectionWorkingGroup:2016ypw} for $m_h=125$~GeV, which has been computed at NNLL: $\sigma^{\rm NNLL}=39.59$~fb, and multiply by the values shown in table~\ref{t-xs}, assuming that most higher order corrections will cancel in the ratio.~\footnote{ Higher order contributions to double Higgs production were calculated in Refs.~\cite{deFlorian:2013uza,deFlorian:2013jea,Maltoni:2014eza,Frederix:2014hta,Grigo:2014jma,Grigo:2015dia,Borowka:2016ehy,Degrassi:2016vss,deFlorian:2016uhr,Spira:2016zna,Borowka:2016ypz,Heinrich:2017kxx,Jones:2017giv,Davies:2018ood,Banerjee:2018lfq,Davies:2018qvx,Baglio:2018lrj,Davies:2019dfy,Chen:2019lzz,Chen:2019fhs,Baglio:2020ini,Muhlleitner:2022ijf,Davies:2022ram,AH:2022elh,Davies:2023vmj,Davies:2023obx,Davies:2023npk,Bi:2023bnq}.}

The analysis of angular variables requires keeping spin correlations in the production as well as in the particle decays. The later ones were included in our simulations by making use of {\tt MadSpin} for the decays of the Higgs to $b\bar b$ and $\gamma\gamma$. We have also included the corrections to the BR($h\to\gamma\gamma$) by the presence of the LQs, as described in sec.~\ref{sec-NPmodel} in the context of the $\kappa$ formalism. The BR($h\to b\bar b$) has been taken from Ref.~\cite{LHCHiggsCrossSectionWorkingGroup:2016ypw}.

\subsection{Backgrounds}
\label{sec-backgrounds}
The backgrounds are given by final states with photons and jets and at detector level one cares about events with two reconstructed photons and two reconstructed $b$-jets. At parton level, the irreducible background without a Higgs is $b\bar b\gamma\gamma$, but $b\bar b\gamma j$ and $b\bar bjj$ with light jets misidentified as photons can also be sizable given the large QCD production cross-section, as well as $c\bar c\gamma\gamma$ and $c\bar c\gamma j$ with $c$-jets misidentified as $b$-jets. Corrections to these backgrounds with one extra final jet are also taken into account. There are in addition backgrounds with a single Higgs, as $Zh$ and $t\bar t h$, for which the two photons reconstruct the Higgs; although these processes have a partonic cross-sections much smaller than other backgrounds, their selection acceptance is much higher~\cite{ATLAS:2017muo}.   

Following the analysis of ATLAS~\cite{ATLAS:2017muo}, we have generated the following backgrounds, that give the leading contributions \footnote{Our selection cuts are quite similar to those implemented in the ATLAS search strategy; the main difference being the $m_{b\bar{b}}$ window. Therefore, we expect the same background hierarchy.}: the irreducible  $b\bar b \gamma\gamma+b\bar b \gamma\gamma j$, the reducible processes $b\bar b \gamma j+b\bar b \gamma jj$ and $c\bar c \gamma\gamma+c\bar c \gamma\gamma j$, the associated production of the Higgs with a top pair, $t\bar t h$, and with a $Z$ boson, $Zh$. For $b\bar b \gamma j+b\bar b \gamma jj$ it is required that one of the light jets produced either at parton level or in the showering and hadronization is misidentified as a photon. The selection of such misidentified jet was carried out by considering a $j\to \gamma$ fake rate of $0.5\times 10^{-3}$ and sampling over all the reconstructed jets. For 
$c\bar c \gamma\gamma+c\bar c \gamma\gamma j$, where one of the $c$-jets must fake a photon, we used the misidentification rate of the default ATLAS HL-LHC card from Delphes, which is roughly of order 10\%, although it depends on its transverse momentum and pseudorapidity. For $t\bar t h$, we took into account both the hadronic and semileptonic decay modes for the tops. Finally, for $Zh$, we included the irreducible contribution from $Z\to b\bar{b}$ and the reducible given by $Z\to c\bar{c}$. Since the misidentification rate of light jets into $b$-jets is significantly smaller than that of $c$-jets , we have not considered this reducible contribution.~\footnote{We have also studied the correction to $Zh$ at one loop by the presence of LQs, this process is initiated by gluons, thus it does not interfere with the LO process in the SM, that is initiated by quarks. We found that the effect in the cross-section is approximately $8\%$ of the LO result for the benchmark point with the lightest LQ, being much smaller for the other benchmark points. As we will show in the next section, the $Zh$ at LO gives a subleading contribution; thus, this NLO correction can be discarded.}
 
\begin{figure}[t!]
\begin{center}
\begin{tabular}{cc}
\hspace*{-9mm}
\includegraphics[width=0.52\textwidth]{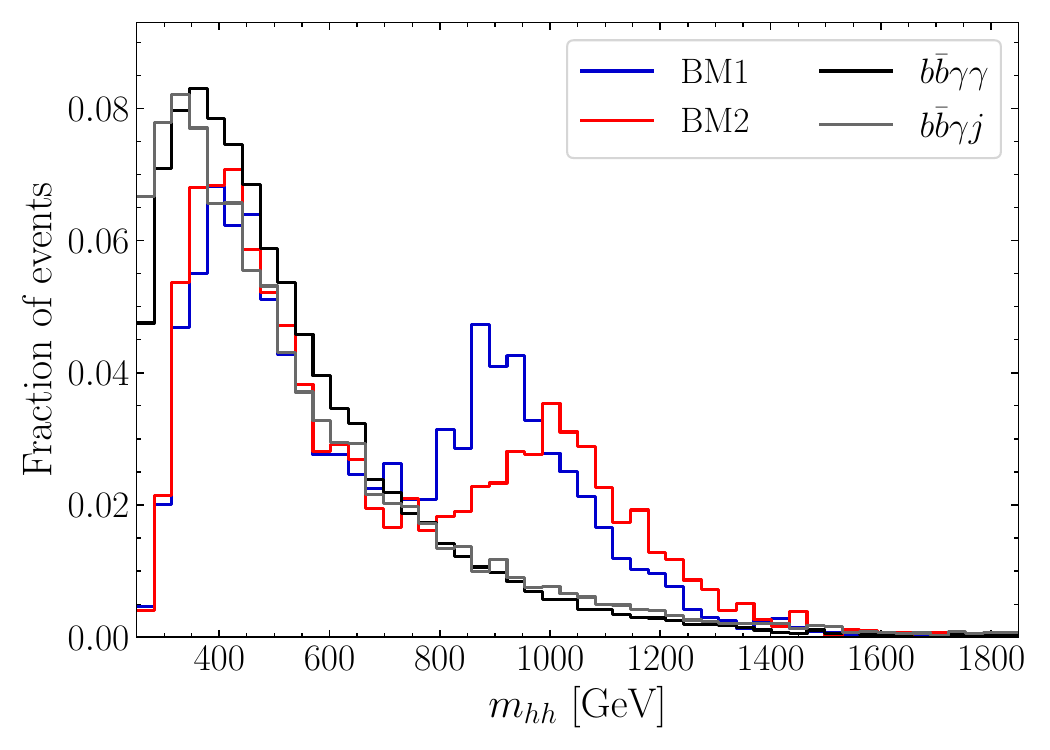} &
\includegraphics[width=0.52\textwidth]{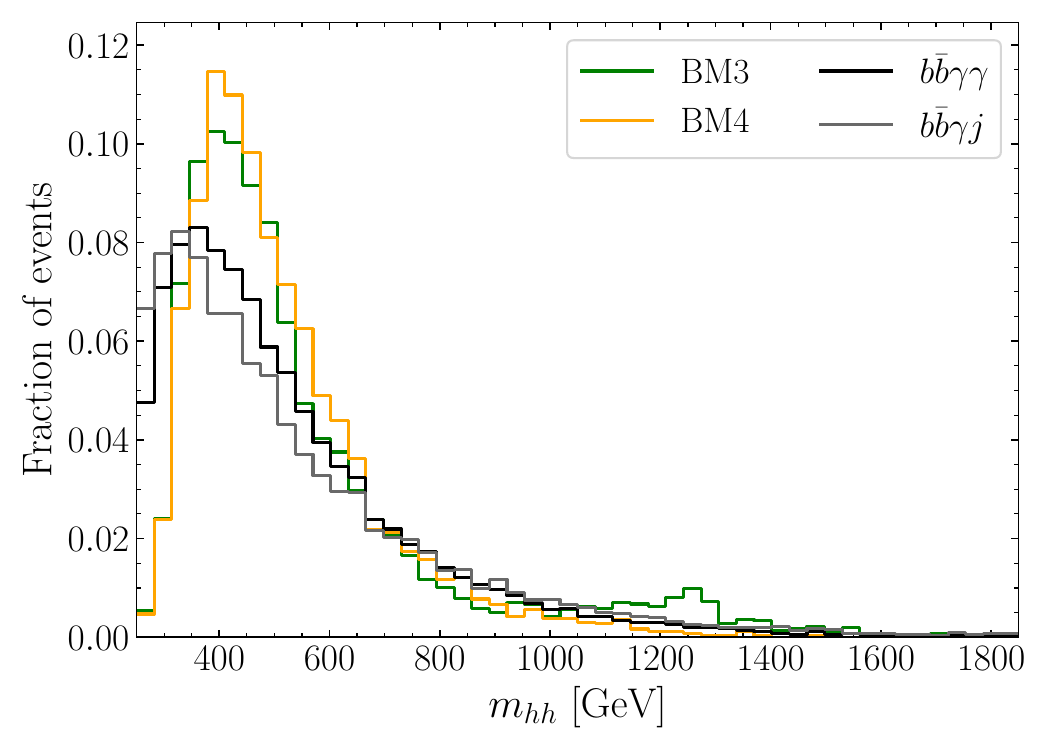}\\
\hspace*{-8mm}
\includegraphics[width=0.52\textwidth]{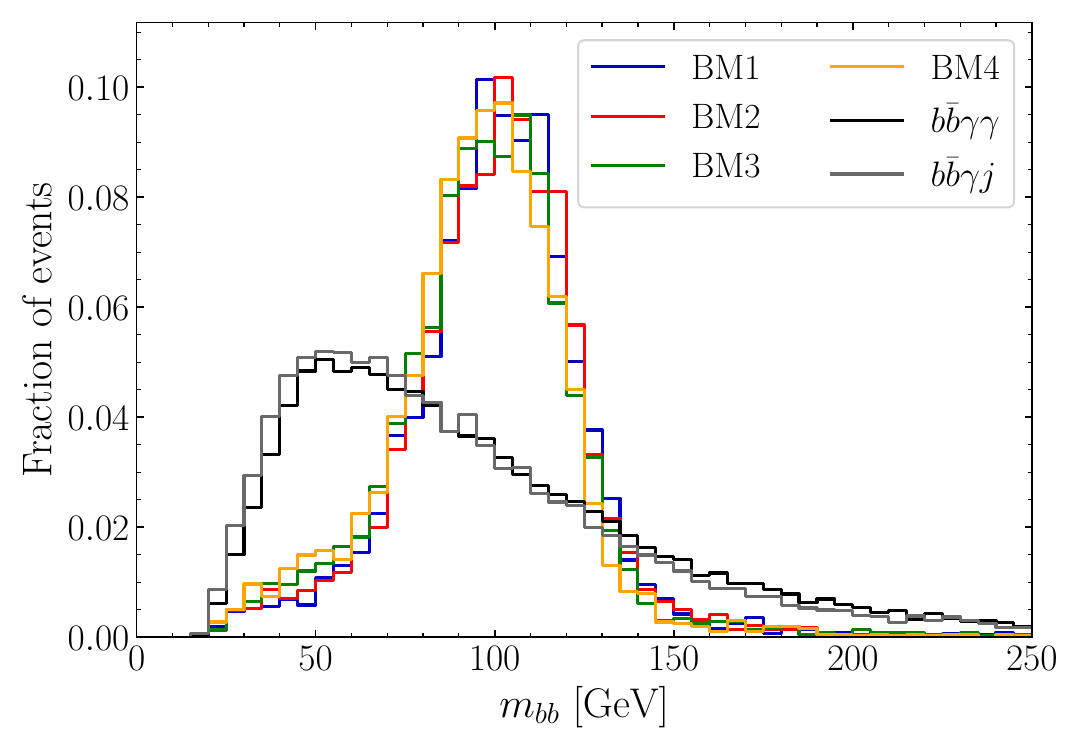} &
\includegraphics[width=0.525\textwidth]{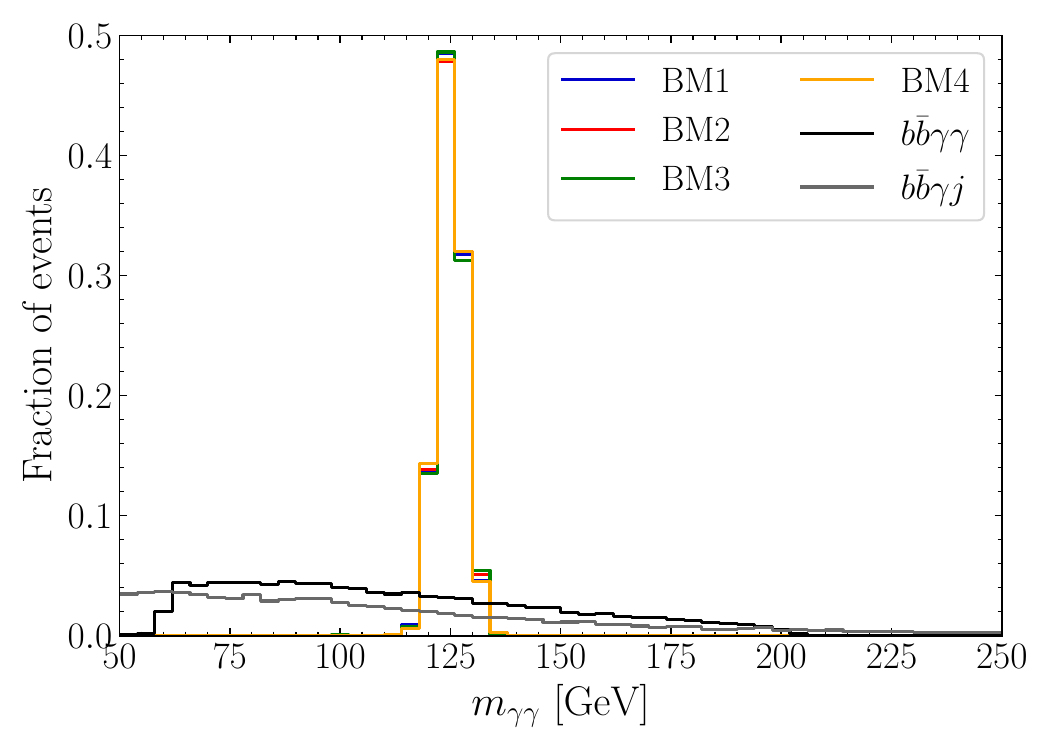}
\end{tabular}
\end{center}
\vspace*{-4mm}
\caption{Invariant mass distributions after applying the selection cuts. Upper panels: $m_{hh}$ distribution for BM1 and BM2 (left panel) and BM3 and BM4 (right panel). Lower panels:  $m_{bb}$ (left panel) and $m_{\gamma\gamma}$ (right panel) distributions. In all the cases the dominant backgrounds $b\bar{b}\gamma\gamma$ and $b\bar{b}\gamma j$ are included for reference.}
\label{fig-invmass-distr}
\end{figure}

\section{Analysis and results}
\label{sec-analysis}
Following the ATLAS analysis in Ref.~\cite{ATLAS:2017muo}, we apply the following selection criteria to the events simulated at detector level:
\begin{itemize}
    \item The number of isolated photons with $p_T >$ 30 GeV and $|\eta|<1.37$ or $1.52<|\eta|<2.37$ is required to be $\geq 2$.
    \item The number of $b$-tagged jets must be $\geq 2$, with the leading and subleading $b$-jets satisfying $|\eta|< 2.4$ and $p_T > 40$ GeV  and 30 GeV, respectively.
    \item Events with more than five jets with $|p_T|>30$ GeV and $|\eta|<2.5$ are discarded. For this requirement, the number of jets in the event includes those tagged as $b$-jets with $|\eta|<2.4$.
    \item Events containing isolated leptons with $p_T> 25$ GeV and $|\eta|<2.5$ are vetoed.
    \item The photon pair candidate to reconstruct one of the Higgs bosons must contain photons that are separated from other photons by $\Delta R_{\gamma\gamma}>0.4$ and from the jets in the event by $\Delta R_{\gamma j}>0.4$. In addition, $\Delta R_{\gamma\gamma}< 2.0$.
    \item The $b$-jets in the pair taken as a candidate to reconstruct one of the Higgs bosons must fulfill $0.4 < \Delta R_{bb}<2.0$.
    \item From all the possible candidate pairs of photons and $b$-jets, those with the closest invariant mass to the Higgs mass are selected. The transverse momenta of the resulting $\gamma\gamma$ and $b\bar{b}$ systems are required to satisfy $p^{\gamma\gamma}_T, p^{bb}_T> 80$ GeV.
\end{itemize}
From now on the above criteria will be denoted collectively as selection cuts. In addition to the cuts listed above, the ATLAS analysis includes the requirements 122 GeV $<m_{\gamma\gamma}<$ 128 GeV and 100 GeV $<m_{bb}<$ 150 GeV. In order to validate our simulation setup we have computed the significance for the SM obtained with the full set of cuts proposed by ATLAS, obtaining 1$\sigma$ for an integrated luminosity of 3 $\mathrm{ab}^{-1}$ which is consistent with the value reported in~\cite{ATLAS:2017muo}.

In Fig.~\ref{fig-invmass-distr} we show the invariant mass distributions $m_{hh}$, $m_{bb}$ and $m_{\gamma\gamma}$ obtained after applying the selection cuts. The four signal benchmarks exhibit peaks around the Higgs mass in the $m_{\gamma\gamma}$ distribution, while in the case of the $m_{bb}$ the peaks are shifted to smaller invariant masses due to the jet energy correction applied to account for differences between the generator and reconstruction levels. The distinctive feature of the LQ model consisting on a second peak around $2\, m_{\chi^u_2}$ in the $m_{hh}$ distribution is also found at detector level and after applying the selection cuts. Moreover, by comparison with Fig.~\ref{fig-mhh-pT12}, we see that the relative proportion of events populating the second peak increases with respect to those in the first peak after imposing the selection cuts due to their bigger impact on events in the region of smaller $m_{hh}$ values. 

\begin{figure}[t!]
\begin{center}
\begin{tabular}{cc}
\hspace*{-7mm}
\includegraphics[width=0.51\textwidth]{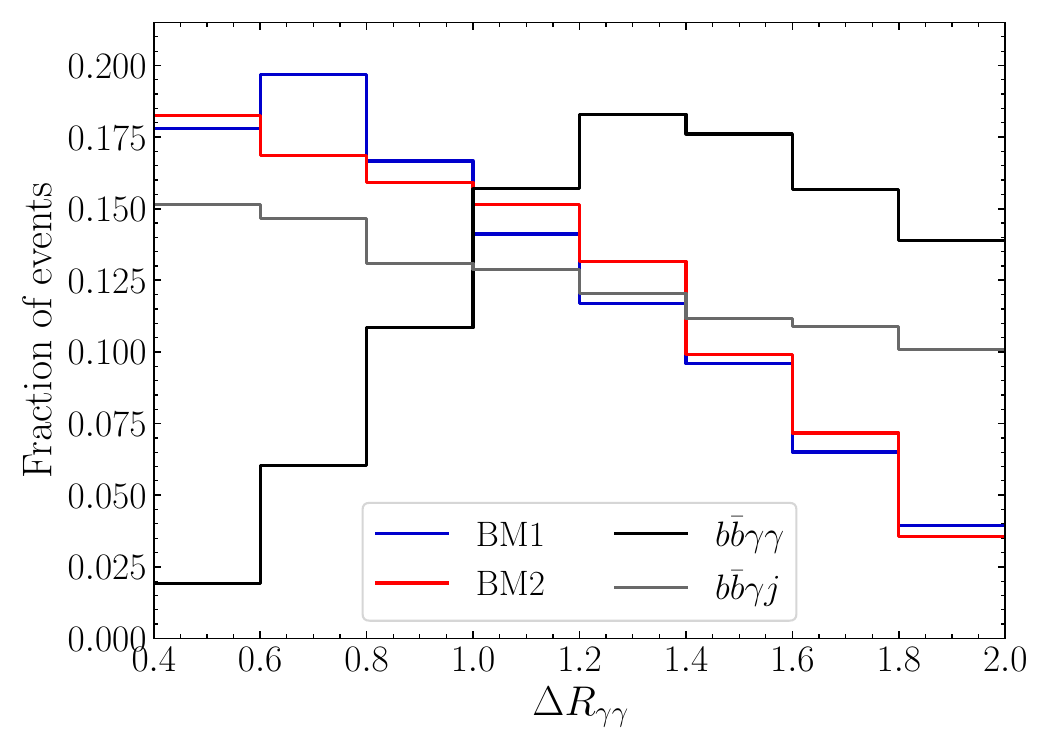} &
\includegraphics[width=0.51\textwidth]{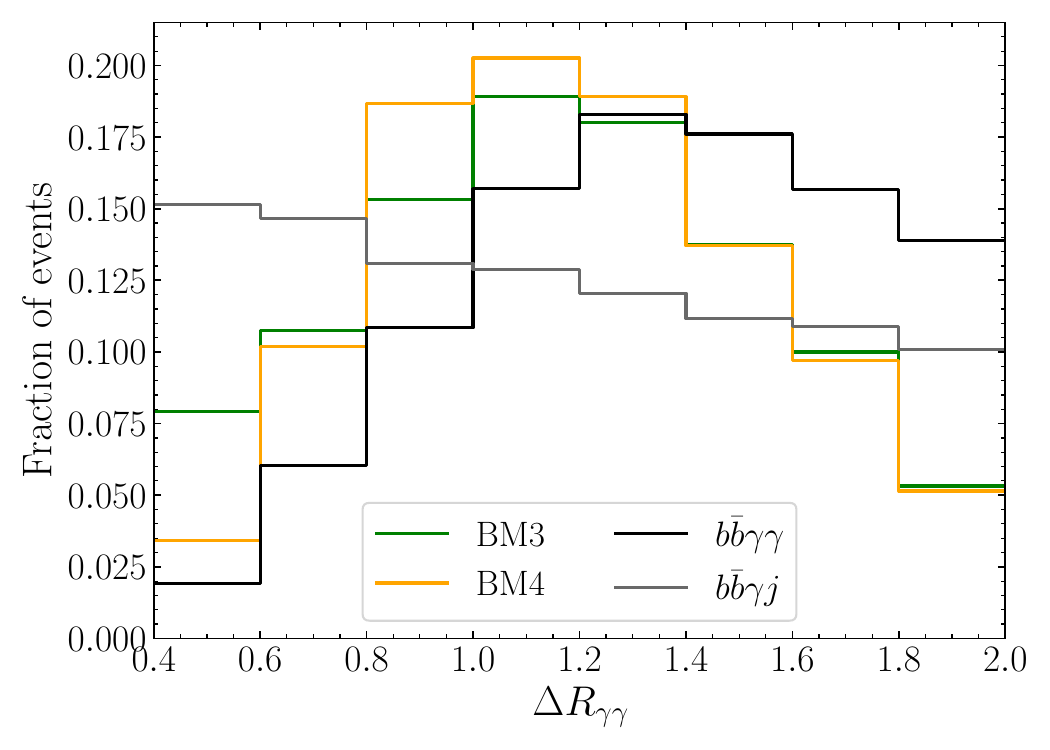}
\end{tabular}
\end{center}
\vspace*{-4mm}
\caption{$\Delta R_{\gamma\gamma}$ distribution for BM1 and BM2 (left panel) and BM3 and BM4 (right panel) after applying the selection cuts. The dominant backgrounds $b\bar{b}\gamma\gamma$ and $b\bar{b}\gamma j$ are included for reference.}
\label{fig-dRaa-full}
\end{figure}

\begin{figure}[t!]
\begin{center}
\hspace*{-8mm}
\begin{tabular}{cc}
\hspace*{-2mm}
\includegraphics[width=0.51\textwidth]{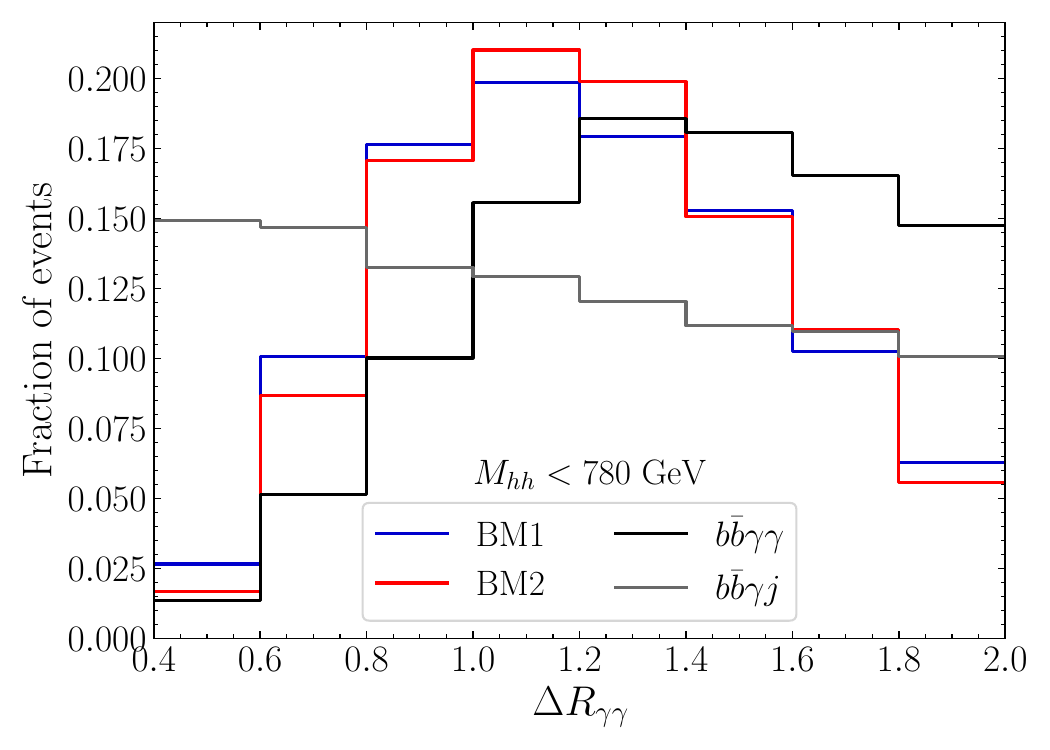} &
\includegraphics[width=0.51\textwidth]{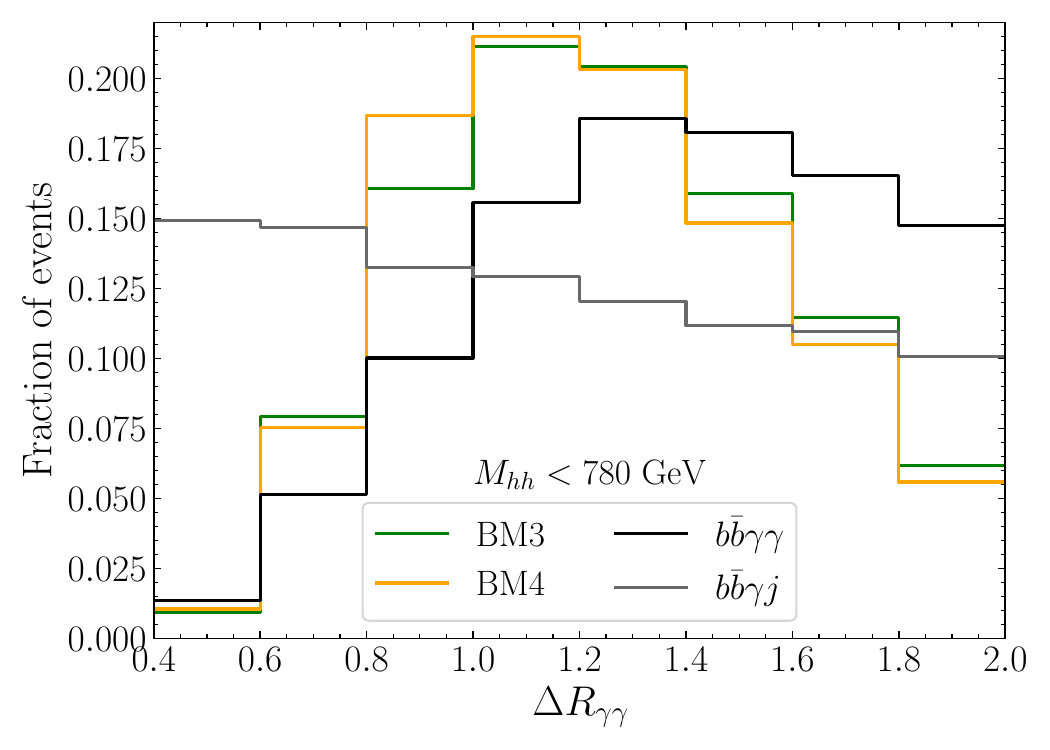}\\
\includegraphics[width=0.51\textwidth]{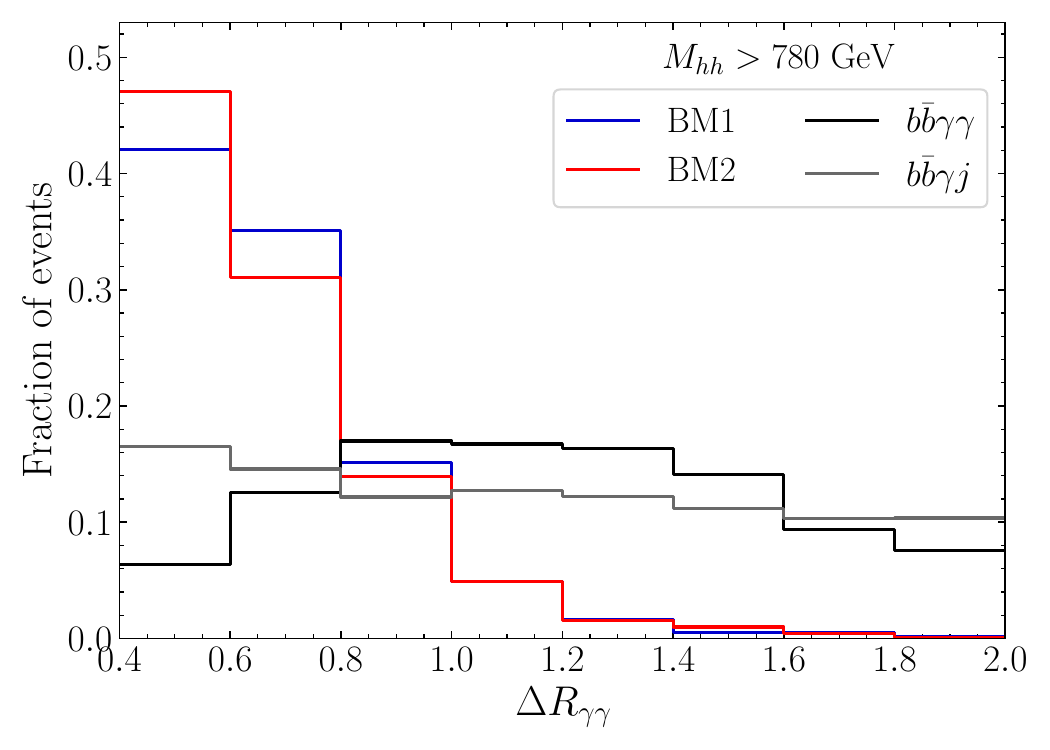} &
\includegraphics[width=0.51\textwidth]{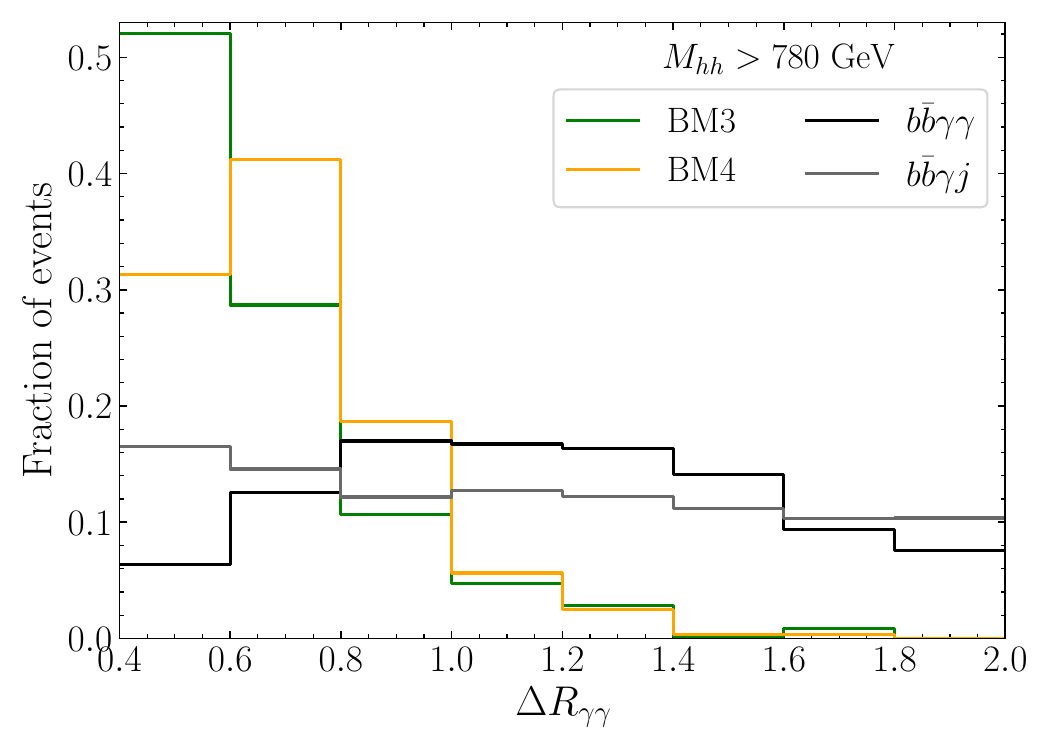}
\end{tabular}
\end{center}
\vspace*{-4mm}
\caption{$\Delta R_{\gamma\gamma}$ distributions after applying the selection cuts and the cut $M_{hh}< 780$ GeV (upper panels) and $M_{hh}> 780$ GeV (lower panels) for BM1 and BM2 (left panels) and BM3 and BM4 (right panels). The dominant backgrounds $b\bar{b}\gamma\gamma$ and $b\bar{b}\gamma j$ are included for reference.}
\label{fig-dRaa-780}
\end{figure}

In Fig.~\ref{fig-dRaa-full} we display the $\Delta R_{\gamma\gamma}$ distribution for the four signal benchmarks. While in the case of BM1 and BM2 the distributions are peaked at small $\Delta R_{\gamma\gamma}$ values, for BM3 and BM4 the peaks are shifted to larger $\Delta R_{\gamma\gamma}$ values. This different behavior can be understood if we separate the contribution of events arising from the two mutually exclusive regions $m_{hh} < 780$ GeV and $>780$ GeV. The corresponding $\Delta R_{\gamma\gamma}$ distributions are shown in Fig.~\ref{fig-dRaa-780}, where we see that for all benchmarks the events with $m_{hh} > 780$ GeV tend to accumulate at small values of $\Delta R_{\gamma\gamma}$ which is expected since for these events the Higgs bosons are more likely to be boosted leading to collimated decay products. Therefore, the shape of the combined distributions in Fig.~\ref{fig-dRaa-full} is driven by the proportion of events in which the Higgs bosons tend to be boosted. In the case of BM1 and BM2 almost 50\% of the events lie in the region $m_{hh} > 780$ GeV, while for BM3 and BM4 only 10\% of the events correspond to this region. The small event yield of these BMs at large $m_{hh}$ is a consequence of the light LQ being heavier ($m_{\chi^u_2}=621$ and 800 GeV, respectively) since in this case the second peak of the distribution is moved to a region where the PDFs are very suppressed.


\begin{figure}[t!]
\begin{center}
\begin{tabular}{cc}
\hspace*{-7mm}
\includegraphics[width=0.51\textwidth]{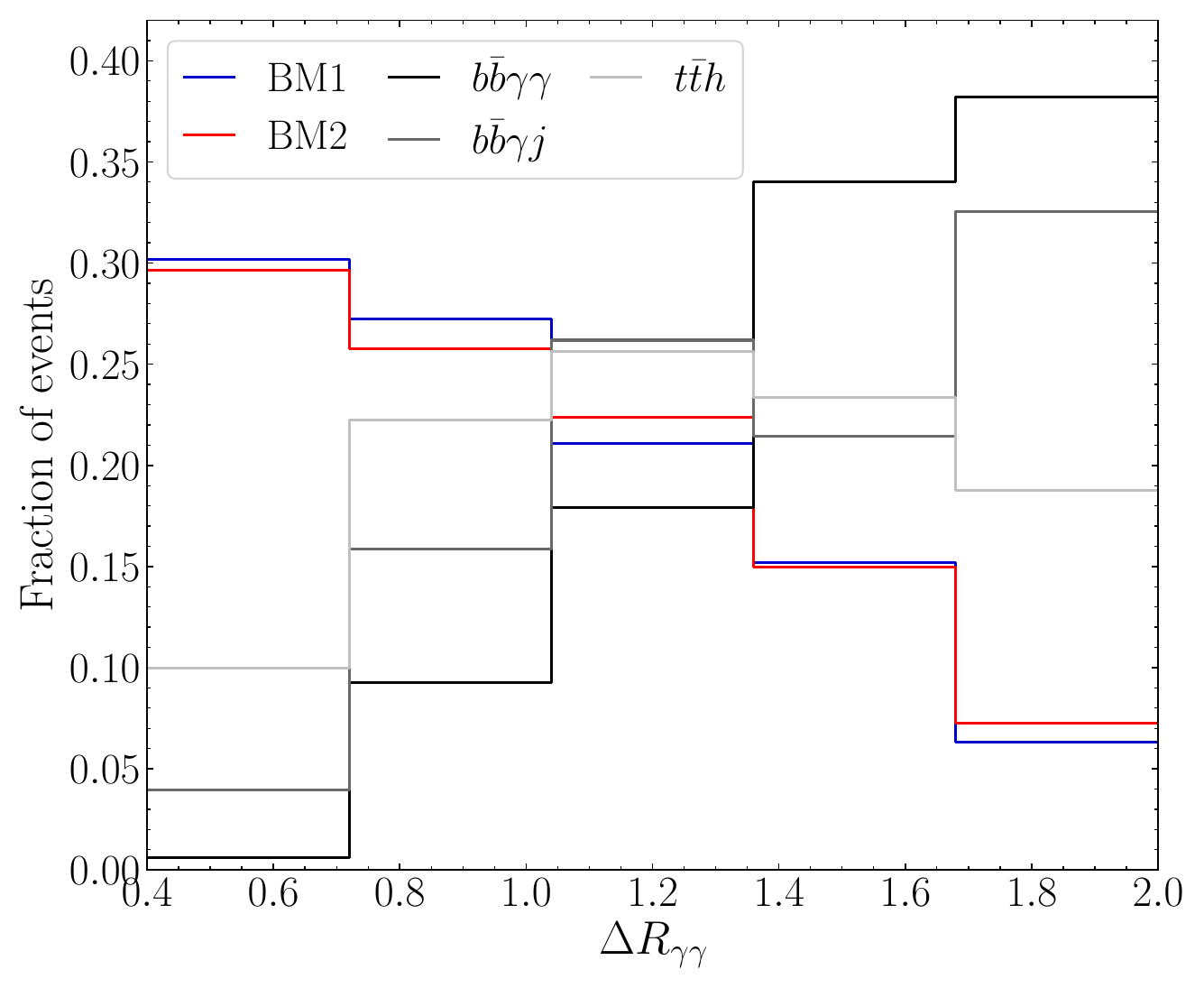} &
\includegraphics[width=0.51\textwidth]{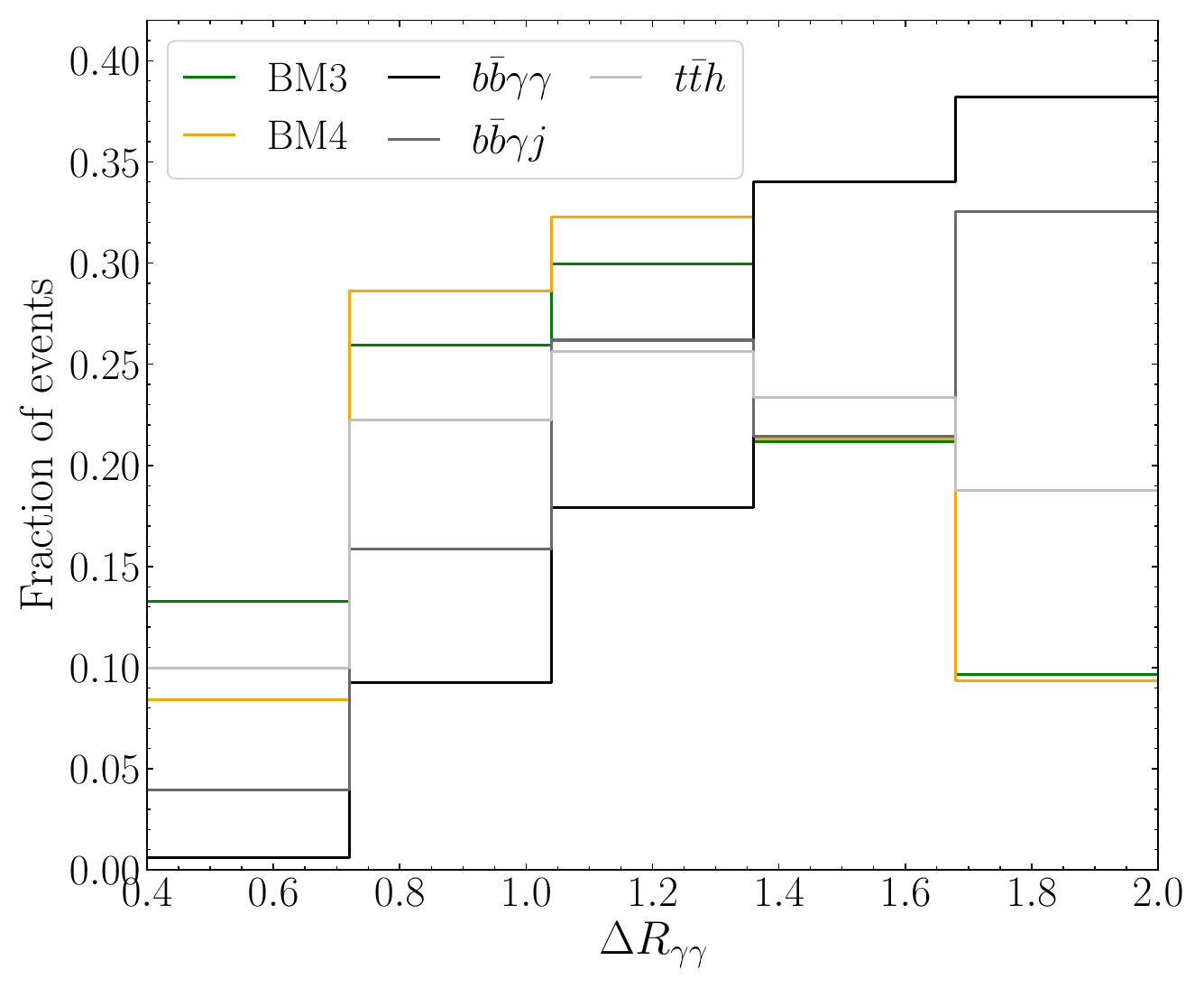}
\end{tabular}
\end{center}
\vspace*{-4mm}
\caption{$\Delta R_{\gamma\gamma}$ distribution for BM1 and BM2 (left panel) and BM3 and BM4 (right panel) after applying the selection cuts along with the invariant mass cuts 122 GeV $< m_{\gamma \gamma}<$ 128 GeV and 80 GeV $< m_{bb}<$ 130 GeV .}
\label{fig-dRaa-5bins}
\end{figure}

Instead of carrying out a global analysis based on cuts, in this work, we take advantage of the discriminating power of the $\Delta R_{\gamma\gamma}$ distribution by performing a statistical analysis based on the binned log-likelihood ratio~\footnote{We have also applied the same statistical analysis discussed in this section to the $\Delta R_{bb}$ distribution. However, since $\Delta R_{\gamma\gamma}$ has proven to be more sensitive and led to better discovery prospects, we exclusively focus this section on the results obtained with it.}. However, for this analysis to be sensitive it is necessary to improve the signal-to-background ratio first by applying further cuts. In particular, we add the same cut on $m_{\gamma\gamma}$ as in the ATLAS analysis but modify the cut on $m_{bb}$ by shifting the mass window to smaller values; specifically, we take 80 GeV $<m_{bb}<$ 130 GeV. The fact that a mass window located at smaller $m_{bb}$ values leads to an improvement in the sensitivity was already pointed out in Ref.~\cite{Adhikary:2017jtu}, even though the cut applied in our case is slightly different. The corresponding cut flow is shown in Table~\ref{tab:cutflow2} for the NP benchmarks and also for the SM along with the main backgrounds. We also include the expected discovery significance computed as~\cite{Cowan:2010js}
\begin{equation}
    \label{eqsig}
    \mathcal{S}=\sqrt{-2\left((s+b)\ln\left(\frac{b}{s+b}\right)+s\right)},
\end{equation}
where $s$ and $b$ are the number of signal and background events, respectively. As expected, the adding of the invariant mass cuts improves the signal-to-background ratio substantially, even to the point of leading to significance values above the evidence level for BM1 and BM2. Moreover, after applying these cuts the $\Delta R_{\gamma\gamma}$ distribution keeps most of the features found in Fig.~\ref{fig-dRaa-full} for the signal, while for the dominant backgrounds the maximum is now reached at larger $\Delta R_{\gamma\gamma}$ values, which increases the discrimination power of this variable (see Fig.~\ref{fig-dRaa-5bins}). We make use of the whole distribution by applying a statistical analysis based on the test statistics $q_0$ and $q_{\mu}$~\cite{Cowan:2010js}, which allows us to determine the expected discovery sensitivity and exclusion limits. The analysis is implemented through the package {\tt pyhf}~\cite{pyhf,pyhf_joss} and the statistical uncertainties are incorporated by modifiers that are paired with constraint terms that limit the rate modification. Specifically, in our analysis, we consider multiplicative modifiers regulated by Poissonian constraint terms.

\renewcommand*{\arraystretch}{1.5}
\begin{table}[t!]
    \small
	\begin{center}
		\begin{tabular}{|c||c|c|c|c|c||c|c|c|c|c|}
\hline
	\textbf{Cut} & BM1 & BM2 & BM3 & BM4 & SM & $b\bar{b}\gamma\gamma$ & $b\bar{b}\gamma j$ & $t\bar{t}h$ & $c\bar{c} \gamma\gamma$ & $Zh$ \\ \hhline{|=||=|=|=|=|=||=|=|=|=|=|}
	Selection cuts & 48.0 & 42.2 & 24.8 & 16.9 & 17.7 & 1178.9 & 1161.2 & 31.6 &86.7 & 8.1 \\ \hline
	$122~\mathrm{GeV} <m_{\gamma\gamma} < 128 ~\mathrm{GeV}$ & 33.3 & 29.2 &17.1 & 11.8 & 12.5 & 56.7 & 31.9 &21.1 & 3.1 & 5.7 \\ \hline
	$80~\mathrm{GeV} < m_{bb} < 130~\mathrm{GeV}$ & 24.9 & 21.7 & 12.4 & 8.6 &9.0 &17.3 & 9.3 & 7.8 & 1.3 & 1.7 \\ \hhline{|=||=|=|=|=|=||=|=|=|=|=|}
    Significance & 3.71 & 3.27 & 1.93 & 1.36 & 1.42 & \multicolumn{5}{c}{}\\
    \cline{1-6}
    \end{tabular}
	\end{center}
	\caption{Cut-flow table listing all the cuts applied prior to the statistical analysis of the $\Delta R_{\gamma\gamma}$ distribution. The displayed event rates correspond to a luminosity of 3 ab$^{-1}$. Significance values computed with Eq.~\ref{eqsig} are also shown.}
	\label{tab:cutflow2}
\end{table}

The results for the discovery sensitivity corresponding to the four benchmarks are summarized in Table~\ref{tab:res}. It is clear that the statistical analysis of the $\Delta R_{\gamma\gamma}$ distribution leads to an improvement of the significance reached by only applying the cuts listed in Table~\ref{tab:cutflow2}. As expected, the increase in the significance is stronger for BM1 and BM2 ($\sim 43\%$) than for BM3 and BM4 ($\sim 21\%$ and $\sim 16\%$, respectively). Finally, for the most promising benchmarks, BM1 and BM2, the evidence level ($3\sigma$) is reached for $905$ and $1180\,\mathrm{fb}^{-1}$, respectively.

In principle, this analysis could also be used to set 95 $\%$ C.L. exclusion limits; in particular, the luminosities required to exclude the model benchmarks 1 and 2 are 600 fb$^{-1}$ and 750 fb$^{-1}$, respectively, while the benchmarks 3 and 4 would require $3\,\mathrm{ab}^{-1}$ and more. However, even if light LQs scenarios such as BM1 an BM2 are not explored anymore by future direct LQs searches, one would still expect these benchmarks to be excluded earlier by single Higgs channels.

We conclude this section providing an estimation of the impact of systematic uncertainties on the background cross section and the reported discovery significances. By enabling the systematics studies in {\tt MadGraph}~\cite{Alwall:2014hca}, we have estimated the uncertainties arising from variations of the factorization and renormalization scales, the scale for the first emission in MLM matching, the merging scale used by {\tt Pythia 8}, and the PDF set, as well as from using different dynamical schemes to set the reference scales. For the three main backgrounds driving the results in Table~\ref{tab:res}, the largest variations in the cross sections are due to the uncertainty in the factorization and renormalization scales and the chosen dynamical scheme, being the impact of the other sources of uncertainty considerably smaller. These uncertainties in the cross section of the backgrounds lead to variations in the expected discovery significances in Table~\ref{tab:res}: for BM1 the central value 5.31 varies within the range 4.79-6.03, while for BM2, with central value 4.68, the range is 4.21-5.33. We see that in both cases the variations are smaller than 14\% and do not change the conclusions regarding the discovery prospects of the most promising benchmarks.

\begin{table}[ht]
\begin{center}
\begin{tabular}{|c|c|c|c|}
\hline
 \rm{BM}1 & \rm{BM}2 & \rm{BM}3 & \rm{BM}4 \\ \hhline{|=|=|=|=|}
 5.31 & 4.68 & 2.33 & 1.62\\
 \hline
\end{tabular}
\end{center}
\caption{Expected discovery significances obtained with the statistical analysis of the $\Delta R_{\gamma\gamma}$ distribution assuming a luminosity of 3 ab$^{-1}$.}
\label{tab:res}
\end{table}

\section{Discussions and conclusions}
\label{sec-conclusions}
We have studied the di-Higgs production initiated by gluon fusion and its decay to $b\bar b\gamma\gamma$ at the HL-LHC, under the presence of new physics with color charge, the main motivation being the possibility of enhancing it due to the presence of new physics. As for the new physics states we considered a minimal set of scalar leptoquarks with the same representations under the SM gauge group as one generation of quarks, such that there are cubic and quartic interactions with the Higgs boson. We selected four benchmark points of the parameter space of the model in which the lightest leptoquark has an electric charge 2/3  and a mass between 400 and 800 GeV, decaying to dijets, leading to benchmarks that are allowed by all current experimental constraints. Studying several differential distributions of $\sigma_{hh}$, first using the numerical code of \cite{DaRold:2021pgn} by analytic calculation of the scattering amplitude, as well as by Monte Carlo simulations via the implementation of the model at one loop level in {\tt MadGraph5}, we find that the presence of the light LQ manifests itself as a resonance in some of the differential distributions. We have also simulated the main SM backgrounds for the final states under consideration. In particular, we looked at differential kinematical distributions like $p_{T,h}$, $m_{hh}$, $m_{pp}$ and $\Delta R_{pp}$, with $p=\gamma$ or $b$. It was found that, after the application of a suitable set of cuts, the distribution $\Delta R_{\gamma\gamma}$ is one of the most sensitive observables for discriminating signal from background, as it is correlated with the presence of the resonant peak associated with the light LQ in the differential distributions.

After applying a set of selection cuts that already improve the signal-to-background ratio, we performed a statistical analysis on the $\Delta R_{\gamma\gamma}$ for the four benchmarks, obtaining significances for $\mathcal{L}=3$ ab$^{-1}$ above the discovery level and close to it for BM1 and BM2 (5.31 and 4.68, respectively). Moreover, we found that the evidence level could be achieved for luminosities of 905 and 1180 fb$^{-1}$, respectively. Regarding the exclusion prospects, luminosities of 600 and 750 fb$^{-1}$ would be enough. The benchmarks 3 and 4 are more challenging since in this case there are very few events with boosted Higgs, and these are precisely the events that make the distribution more efficient to discriminate the signal from the backgrounds. The significances reached for BM3 and BM4 are in fact below the evidence level: 2.33 and 1.62, respectively. For these benchmarks, one may wonder if a strategy incorporating machine-learning tools could improve the discovery prospects. 

Deep neural networks (DNNs) are a well-known example of these powerful algorithms. DNNs can be trained to perform non-linear global cuts that make use of complex and non-intuitive patterns to improve human performance in classification tasks. As part of an ongoing work, we carried out a preliminary study in which we optimized DNN models from statistically balanced samples of signal and background events tagged by the same five kinematical features considered in the analysis of Section \ref{sec-analysis}: $m_{bb}$, $m_{\gamma\gamma}$, $m_{hh}$, $\Delta R_{bb}$ and $\Delta R_{\gamma\gamma}$. We reached discovery significances of $2.65$ for BM3 and $1.69$ for BM4, which means an improvement compared to Table \ref{tab:cutflow2}. As an example of the impact of potential background uncertainties, we found that the significances decrease to $2.25$ and $1.31$ when a $15\%$ of systematic uncertainty is considered. These results are below those shown in Table \ref{tab:res} for the binned log-likelihood analysis. Further tests, including additional kinematical variables in the training phase and different DNNs architectures will be presented in a future article.

To conclude, a similar analysis based on kinematical distributions could be applied to other models with loop-enhanced di-Higgs production at the LHC. In particular, it would be very interesting to establish the reach with new colored fermions running in the loop. We leave this study for a future work.

\section*{Acknowledgments}
This work has been partially supported by CONICET and ANPCyT
projects PIP-11220200101426 and PICT-2018-03682. A.M. thanks in particular Prof. Carlos E. M. Wagner and the hospitality from the EFI at the University of Chicago and the HEP division at ANL where part of this work was finished. 

\bibliography{biblio}{}
\bibliographystyle{JHEP}

\end{document}